\documentclass{gtart}

\usepackage{amsmath}
\usepackage{amssymb}
\usepackage{verbatim}
\usepackage{epsf}

\input gtoutput
\volumenumber{3}\papernumber{2}\volumeyear{1999}
\pagenumbers{21}{66}\published{22 March 1999}
\proposed{Robion Kirby}\seconded{Vaughan Jones, Walter Neumann}
\received{9 January 1999}
\accepted{9 March 1999}

\theoremstyle{plain}
\newtheorem{theorem}{Theorem}
 
\newtheorem{lemma}[theorem]{Lemma}
\newtheorem{coroll}[theorem]{Corollary}

\theoremstyle{definition}
\newtheorem{defn}[theorem]{Definition}
\newtheorem*{rem}{Remark}

\newtheorem*{ack}{Acknowledgements}

\newtheorem*{que}{Question}
\newcommand{\bqu}{\tiny\begin{que}}
\newcommand{\equ}{\end{que}\normalsize}

\DeclareMathOperator{\Hess}{Hess}
\DeclareMathOperator{\vol}{vol}
\DeclareMathOperator{\area}{vol}
\DeclareMathOperator{\sgn}{sgn}

\newcommand{\C}{\mathbb C}
\newcommand{\R}{\mathbb R}
\newcommand{\Z}{\mathbb Z}
\newcommand{\Q}{\mathbb Q}
\newcommand{\N}{\mathbb N}
\newcommand{\PP}{\mathbb P}

\newcommand{\Rtwo}{{\mathbb R}^2}
\newcommand{\Rthree}{{\mathbb R}^3}

\newcommand{\ra}{\rightarrow}
\newcommand{\hra}{\hookrightarrow}
\newcommand{\lan}{\langle}
\newcommand{\ran}{\rangle}
\newcommand{\norm}{\|}

\newcommand{\LL}{\mathcal L}
\newcommand{\GC}{G^{\mathbb C}}

\newcommand{\lie}[1]{\mathfrak{#1}}
\newcommand{\coo}[1]{\mathfrak{#1}^*}
\newcommand{\clie}[1]{\mathfrak{#1}^{\mathbb C}}

\newcommand{\sixj}{\left\{ \begin{matrix} a&b&c\\ d&e&f
\end{matrix} \right\} }
\newcommand{\ksixj}{\left\{ \begin{matrix} ka&kb&kc\\ kd&ke&kf
\end{matrix} \right\} }
\newcommand{\sixjj}{\left\{ \begin{matrix} a'&b'&c'\\ d'&e'&f'
\end{matrix} \right\} }

\newcommand{\e}[2]{\alpha_{#1#2}}
\newcommand{\Pol}{\mathcal P}

\begin{document}

\title{Classical $6j$--symbols and the tetrahedron} 
\asciititle{Classical 6j-symbols and the tetrahedron}

\author{Justin Roberts} 

\address{Department of Mathematics and Statistics\\Edinburgh 
University, EH3 9JZ, Scotland}
\email{justin@maths.ed.ac.uk}

\begin{abstract}
A classical $6j$--symbol is a real number which can be associated to a
labelling of the six edges of a tetrahedron by irreducible
representations of $SU(2)$. This abstract association is traditionally
used simply to express the symmetry of the $6j$--symbol, which is a
purely algebraic object; however, it has a deeper geometric
significance. Ponzano and Regge, expanding on work of Wigner, gave a
striking (but unproved) asymptotic formula relating the value of the
$6j$--symbol, when the dimensions of the representations are large, to
the volume of an honest Euclidean tetrahedron whose edge lengths are
these dimensions. The goal of this paper is to prove and explain this
formula by using geometric quantization. A surprising spin-off is that
a generic Euclidean tetrahedron gives rise to a family of twelve
scissors-congruent but non-congruent tetrahedra.
\end{abstract}
\asciiabstract{%
A classical 6j-symbol is a real number which can be associated to a
labelling of the six edges of a tetrahedron by irreducible
representations of SU(2). This abstract association is traditionally
used simply to express the symmetry of the 6j-symbol, which is a
purely algebraic object; however, it has a deeper geometric
significance. Ponzano and Regge, expanding on work of Wigner, gave a
striking (but unproved) asymptotic formula relating the value of the
6j-symbol, when the dimensions of the representations are large, to
the volume of an honest Euclidean tetrahedron whose edge lengths are
these dimensions. The goal of this paper is to prove and explain this
formula by using geometric quantization. A surprising spin-off is that
a generic Euclidean tetrahedron gives rise to a family of twelve
scissors-congruent but non-congruent tetrahedra.}

\primaryclass{22E99}                
\secondaryclass{81R05, 51M20}              
\keywords{$6j$--symbol, asymptotics, tetrahedron, Ponzano--Regge formula,
 geometric quantization, scissors congruence}          
\asciikeywords{6j-symbol, asymptotics, tetrahedron, Ponzano-Regge formula,
 geometric quantization, scissors congruence}

\maketitlepage



\section{Introduction}

A classical $6j$--symbol is a real number which can be associated to a
labelling of the six edges of a tetrahedron by irreducible
representations of $SU(2)$, in other words by natural numbers. Its
definition is roughly as follows.

Let $V_a$ ($a=0,1,2,\ldots$) denote the $(a+1)$--dimensional
irreducible representation. The $SU(2)$--invariant part of the triple
tensor product $V_a \otimes V_b \otimes V_c$ is non-zero if and only
if
\begin{equation}\label{e:triang} a\leq b+c\qquad  b\leq c+a \qquad c\leq a+b \qquad a+b+c\qua \hbox{is even}\end{equation}
in which case we may pick, almost canonically, a basis vector
$\epsilon^{abc}$ (details are given below).

\begin{figure}[h]\small
\begin{equation*}\vcenter{\hbox{\mbox{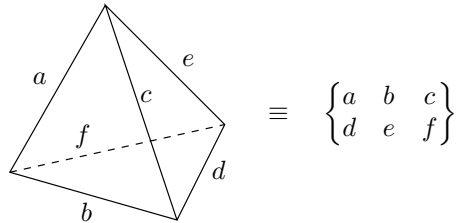}}}\quad \equiv  
\quad \sixj \end{equation*}
\caption{Pictorial representation\label{figtet}}
\end{figure}

Suppose we have a tetrahedron, labelled so that the three labels
around each face satisfy these conditions: we will call this an {\em
admissible} labelling. Then we may associate to each face an
$\epsilon$--tensor, and contract these four tensors together to obtain
a scalar, the $6j$--symbol, denoted by a picture or a bracket symbol as
in figure \ref{figtet}.

This tetrahedral picture is traditionally used simply to express the
{\em symmetry} of the $6j$--symbol, which is naturally invariant under
the full tetrahedral group $S_4$. However, it has a deeper {\em
geometric} significance. To an admissibly-labelled tetrahedron we may
associate a metric tetrahedron $\tau$ whose side lengths are the six
numbers $a,b,\ldots,f$. Its individual faces may be realised in
Euclidean $2$--space, by the admissibility condition
\eqref{e:triang}. As a whole, $\tau$ is either {\em Euclidean}, {\em
Minkowskian} or {\em flat} (in other words has either a non-degenerate
isometric embedding in Euclidean or Minkowskian $3$--space, or has an
isometric embedding in Euclidean $2$--space), according to the sign of
a certain polynomial in its edge-lengths. If $\tau$ is Euclidean, let
$\theta_a, \theta_b, \ldots, \theta_f$ be its corresponding exterior
dihedral angles and $V$ be its volume.

\begin{theorem}[Asymptotic formula]\label{t:main}
Suppose a tetrahedron is admissibly labelled by the numbers
$a,b,c,d,e,f$. Let $k$ be a natural number. As $k \ra \infty$, there
is an asymptotic formula
\begin{equation}\label{e:form} \ksixj \sim \begin{cases}{\displaystyle\sqrt{\frac{2}{3\pi Vk^3}}\cos{\left\{ \sum (ka+1)
\frac{\theta_a}{2} + \frac{\pi}{4}\right\}}} &\hbox{{if $\tau$ is
Euclidean}} \\ \hbox{{exponentially decaying}}&\hbox{{if $\tau$ is
Minkowskian}}\end{cases} \end{equation}
{\rm (}where the sum is over the six edges of the tetrahedron{\rm )}.
\end{theorem}
A (slightly different) version of this formula was conjectured in 1968
by the physicists Ponzano and Regge, building on heuristic work of
Wigner; they produced much evidence to support it but did not prove
it. It is the purpose of this paper to prove the above theorem using
geometric quantization, and to explain the relation between $SU(2)$
representation theory and the geometry of $\Rthree$.

The formula has a lovely and peculiar consequence in elementary
geometry. It is well-known that a generic tetrahedron is not {\em
congruent} (by an orientation-preserving isometry of $\Rthree$) to its
mirror-image, but is {\em scissors-congruent} to it (in other words,
the two tetrahedra are finitely equidecomposable). Inspired by the
additional algebraic {\em Regge symmetry} of $6j$--symbols and the
asymptotic formula above, one may derive from a generic tetrahedron a
family of {\em twelve} non-congruent but scissors-congruent
tetrahedra!

Section 2 contains the algebraic and section 3 the
differential-geometric preliminaries. Section 4 is a warm-up example,
computing asymptotic rotation matrix elements for $SU(2)$
representations. It works in the same way as the eventual computation
(in section 5) for the $6j$--symbol, but is much simpler and displays
the method more clearly. Section 6 contains the geometric corollaries
mentioned above and further notes on the Ponzano--Regge paper.

Throughout the paper, the symbol ``$\sim$'' denotes an asymptotic formula,
whereas ``$\approx$''denotes merely an approximation.

\begin{ack}
After having the basic ideas for this paper, I spent some time
collaborating with John Barrett, trying to find a good method of doing
the actual calculations required. Neither of us had much success
during this period, and the details presented here were worked out by
me later. (I feel quite embarrassed at ending up the sole author in
this way.) I am especially grateful to John for many lengthy,
interesting and helpful discussions on the subject, and also to J\o rgen
Andersen, Johan Dupont, James Flude, Elmer Rees, Mike Singer and
Vladimir Turaev for other valuable discussions.
\end{ack}


\section{Definition and interpretation of $6j$--symbols}

\subsection{Combinatorial definition}

\begin{figure}[ht]
\begin{equation*} \vcenter{\hbox{\mbox{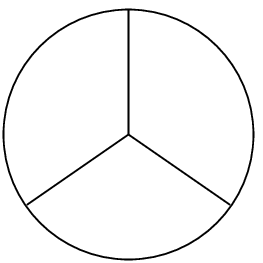}}} \qquad \qquad
\vcenter{\hbox{\mbox{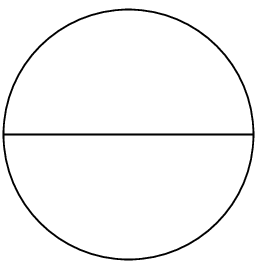}}}\end{equation*}
\caption{Mercedes and theta graphs\label{fskein}}
\end{figure}

The simplest definition is via Penrose's spin network calculus, which
is related to Kauffman bracket skein theory at $A=\pm 1$. The details
are in the book of Kauffman and Lins \cite{KL}. There is a topological
invariant $\langle\,\rangle$ of planar links (systems of generically immersed
curves) defined by sending a link $L$ to
\begin{equation}\label{e:spinor}\langle L\rangle =(-2)^{\text{\em number of
loops{\rm (}L{\rm )}}}(-1)^{\text{\em number of crossings{\rm (}L{\rm )}}}.\end{equation} It
extends to an invariant of suitably-labelled trivalent graphs in
$S^2$, for example the Mercedes (tetrahedron) and theta symbols shown
in figure \ref{fskein}.
To define it, we replace each edge by a number of parallel strands
equal to its label, and connect them up without crossings at the
vertices (this imposes precisely the conditions \eqref{e:triang} on the
the three incident labels). Then we replace this diagram by the set of
all planar links obtained by inserting a permutation of the strands
near the middle of each edge. Finally, evaluate each of these using
\eqref{e:spinor}, add up their contributions, and divide by the number
of such diagrams (the product of the factorials of the edge-labels).
Explicit evaluations of these quantities are given in \cite{KL}. 

\begin{defn}\label{combdef}
The $6j$--symbol shown in figure \ref{figtet} is defined as the
spin-network evaluation of the above admissibly-labelled Mercedes
symbol, divided by the product of the square-roots of the absolute
values of the four theta symbols associated with its vertices. It is
manifestly $S_4$--invariant.
\end{defn}

\begin{rem}
It is important to note that the spin-network picture is {\em dual} to
the one drawn in figure \ref{figtet}. There, the trilinear
invariant spaces are associated with {\em faces} of the tetrahedron,
whereas in the Mercedes symbol they are associated with {\em
vertices}.
\end{rem}

Although this definition is the simplest, we will need a more
algebraic version where the $6j$--symbol is exhibited as a hermitian
pairing of two vectors.

\subsection{Algebraic definition}\label{4def}

Let $V_1$ be the fundamental $2$--dimensional representation of
$SU(2)$, which we will consider as the space of linear homogeneous
polynomials in coordinate functions $Z$ and $W$. Then the other
irreducibles, the symmetric powers $V_a = S^aV$, $a=0, 1, 2, \ldots$,
are the spaces of homogeneous polynomials of degree $a$.  The
dimension of $V_a$ is $a+1$; when $a$ is even, it is an irreducible
representation of $SO(3)$.

Making $Z$, $W$ orthonormal determines an invariant hermitian inner
product $( - , - )$ on $V_1$, and induces inner products on the higher
representations $V_a$, thought of as subspaces of the tensor powers of
$V_1$. The fundamental representation has an invariant skew tensor $Z
\otimes W - W \otimes Z$, which induces quaternionic or real
structures on the $V_a$, according as $a$ is odd or even.

The $SU(2)$--invariant part of the tensor product of two irreducibles
$V_a \otimes V_b$ is zero unless $a=b$, when it is one-dimensional.
Similarly, the invariant part of the triple tensor product $V_a
\otimes V_b \otimes V_c$ of irreducibles is either empty or
one-dimensional, according to the famous conditions \eqref{e:triang}.
(The meaning of the parity condition is clear from the fact that the
centre of $SU(2)$ is the cyclic group $\Z_2$. The other conditions,
often written more compactly as $\vert a-b \vert \leq c \leq a+b$, are
more surprising. Why the existence of a Euclidean triangle with the
prescribed sides should have anything to do with this will be
explained shortly.)

We want to pick well-defined basis vectors $\epsilon^{aa}$ and
$\epsilon^{abc}$ for these spaces of bilinear and trilinear
invariants. Since each such space has a hermitian form and a real
structure, we could just pick real unit vectors, but this would still
leave a sign ambiguity. To fix this we may as well just write down the
invariants concerned. Consider the vectors corresponding to the
polynomials
\[  (Z_1 W_2 - W_1 Z_2)^a \qquad (Z_1 W_2 - W_1 Z_2)^k(Z_1 W_3 - W_1 Z_3)^j(Z_2 W_3 -
W_2 Z_3)^i \] on $\C^2\oplus \C^2$ and $\C^2\oplus \C^2\oplus \C^2$
respectively, where $i=(b+c-a)/2$, $j=(a+c-b)/2$, $k=(a+b-c)/2$. The
required vectors are obtained from these by rescaling using positive
real numbers, to obtain $\epsilon^{aa}$ with norm $\sqrt{a+1}$ and
$\epsilon^{abc}$ with norm $1$.

\begin{defn}\label{6jdef}
Given six irreducibles $V_a, V_b, \ldots, V_f$, one can form
$\epsilon^{abc} \otimes \epsilon^{cde} \otimes \epsilon^{efa} \otimes
\epsilon^{fdb}$ (supposing these all exist) inside a 12--fold tensor
product of irreducibles. One may always form $\epsilon^{aa} \otimes
\epsilon^{bb} \otimes \cdots \otimes \epsilon^{ff}$, and permute the
factors (without reversing the order of the paired factors) to
match. Then the hermitian pairing of these two vectors (inside the
12--fold tensor product) defines the associated $6j$--symbol by 
\[ \sixj = (-1)^{\sum a} ( \overset{6}{\otimes} \epsilon^{aa},
\overset{4}{\otimes} \epsilon^{abc}) \] where $\sum a$ is simply the
sum of the six labels. One should think of it as a function of six
natural numbers $a,b, \ldots, f$, defined whenever the triples
$(a,b,c), (c,d,e), (e,f,a), (f,d,b)$ satisfy the triangle and parity
conditions \eqref{e:triang}, in other words when the associated
tetrahedron labelling is admissible. It is standard to extend the
definition to all such sextuples by setting the $6j$--symbol to zero
elsewhere.
\end{defn}

\begin{lemma}
These two definitions agree.
\end{lemma}
\begin{proof}[Proof {\rm (}Sketch{\rm )}] The Mercedes spin network evaluation used in definition
\ref{combdef} can be reinterpreted as an explicit tensor contraction,
using Penrose's diagrammatic tensor calculus (see \cite{KL}). The
invariant $\langle L\rangle$ of a planar link may be evaluated by making the link
Morse with respect to the vertical axis in $\Rtwo$, replacing cups and
caps with $i$ times the standard skew tensor ($Z \otimes W - W \otimes
Z \in \C \otimes \C$ and its dual, crossings with the flip tensor, and
composing these morphisms to obtain a scalar. If we draw the Mercedes
graph as in figure \ref{messy} and use this recipe to compute it, we
see that it is given as the composition of a vector in $V_1^{\otimes 2
\sum a}$ (coming from the cups in the lower half of the diagram), a
tensor product of twelve Young symmetrisers (coming from the flips
associated to the crossings which are introduced where the labels are)
and a vector in the dual of $V_1^{\otimes 2 \sum a}$ (coming from the
caps).
\begin{figure}[ht]
\[ \vcenter{\hbox{\mbox{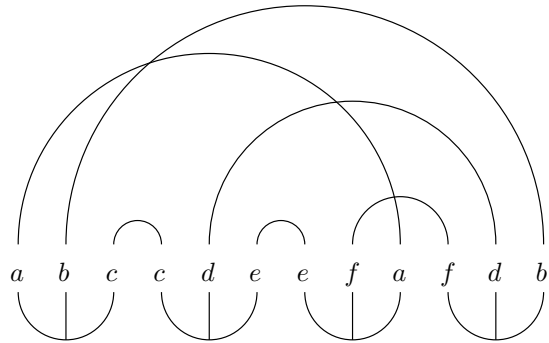}}}  \]
\caption{Writing the $6j$--symbol as a pairing\label{messy}}
\end{figure}
This can be interpreted as a bilinear pairing between a vector in the
tensor product of twelve irreps and one in its dual, if we use the
symmetrisers to project to these. Reinterpreting it as a hermitian
pairing and including the normalisation factors gives the purely
algebraic formulation of definition \ref{6jdef}.
\end{proof}

\begin{rem} This definition does not depend on the choice of hermitian form,
coordinates, or real structure. It does depend on the sign
conventions, but these can be seen to be sensible (in that the
resulting $6j$--symbol is $S_4$--invariant) using the lemma.
\end{rem}

\begin{rem}
A third way of defining the $6j$--symbols is to build a basis for $(V_a
\otimes V_b \otimes V_c \otimes V_d)^{SU(2)}$ out of the trilinear
invariants using an isomorphism such as \[ (V_a \otimes V_b \otimes
V_c \otimes V_d)^{SU(2)} \cong \bigoplus_e (V_a \otimes V_b \otimes
V_e)^{SU(2)} \otimes (V_e \otimes V_c \otimes V_d)^{SU(2)}\] where
$e$ runs through all values such that $(a,b,e)$ and $(e,c,d)$ satisfy
\eqref{e:triang}. There are three standard ways of doing this,
corresponding to the three pairings of the four ``things'' $a,b,c,d$,
and the change-of-basis matrix elements are (after mild
renormalisation) the $6j$--symbols. Using this definition makes the
Elliott--Biedenharn identity (pentagon identity) for $6j$--symbols very
clear, but disguises their tetrahedral symmetry; therefore we will not
consider this method here. See Varshalovich et al \cite{V} for this
approach. Their definition coincides with the two given here, and with
the one in Ponzano and Regge (though in these physics-oriented papers,
half-integer spins are used).
\end{rem}

\subsection{Heuristic interpretation}

The representation theory of $SU(2)$ is well-known to physicists as
the theory of quantized angular momentum. The fundamental
$2$--dimensional complex representation $V_1$ can be viewed as the
space of states of spin of a spin--$\frac12$ particle. The other
irreducibles, the symmetric powers $\{V_a=S^aV, a \in \N \}$, are
state spaces for particles of higher spin; indeed, physicists label
them by their associated spins $j=\frac12a$. Quantum and classical
state-spaces are very different: the classical state of a spinning
particle is described by an angular momentum vector in $\Rthree$,
whereas in the quantum theory, one should imagine the state vectors as
wave-functions on $\Rthree$, whose pointwise norms give {\em
probability distributions} for the value of the angular momentum
vector. However, when the spin is very large, the quantum and
classical behaviour should begin to correspond. For example, the
wave-functions representing states of a particle with large spin $j$
should be concentrated near the sphere of radius $j$ in $\Rthree$.

Many representation-theoretic quantities, most obviously square-norms
of matrix elements of representations, can be interpreted as
probability amplitudes for quantum-mechanical observations. Wigner
\cite{Wig} explained the $6j$--symbol as follows. Suppose one has a
system of four particles with spins $\frac12a, \frac12b, \frac12c,
\frac12d$ and total spin $0$. Then the {\em square} of the $6j$--symbol
is essentially the probability, given that the first two particles
have total spin $\frac12e$, that the first and third combined have
total spin $\frac12f$. (Compare with the third definition in the
remark in subsection \ref{4def}.) He
reasoned that for large spins, because of the concentration of the
wave-functions, one can treat this statement as dealing with addition
of vectors in $\Rthree$. Suppose one has four vectors of lengths
$\frac12a, \frac12b, \frac12c, \frac12d$ which form a closed
quadrilateral. Then, given that one diagonal is $\frac12e$, what is
the probability that the other is $\frac12f$? His analysis yielded
the formula:
\[ \sixj ^2 \approx \frac{1}{3 \pi V} \]
where $V$ is the volume of the Euclidean tetrahedron whose
edge-lengths are $a, b, \ldots, f$, supposing it exists. He emphasised
that this is a dishonest approximation: the $6j$--symbols are wildly
oscillatory functions of the dimensions, and his formula is only a
local average over these oscillations, true in the same sense that one
might write:
\[ \hbox{For $\theta \neq 0$,}\ \cos^2(k\theta) \approx \frac12 \quad
\hbox{as $k \ra \infty$}.\] 

Ponzano and Regge improved his formula to one very similar to
\eqref{e:form}, deducing the oscillating phase term from clever
empirical analyses, and verified that as an approximation it is
extremely accurate, even for small irreps.


\section{Geometric quantization}

\subsection{Borel--Weil--Bott}
To rigorize Wigner's arguments, we need a concrete geometric
realisation of the representations $V_a$. This is provided by the
Borel--Weil--Bott theorem (see for example Segal \cite{CSM} or
\cite{FH}): all finite-dimensional irreducible representations of
semisimple Lie groups are realised as spaces of holomorphic sections
of line bundles on compact complex manifolds, on which the groups act
equivariantly. We only need the simplest case of this, namely that the
irrep $V_a$ of $SU(2)$ is the space of holomorphic sections of the
$a$th tensor power of the hyperplane bundle $\LL$ on the Riemann
sphere $\PP^1$. If one thinks of these as functions on the dual
tautological bundle, which is really just $\C^2$ blown up at the
origin, they can be identified with spaces of homogeneous polynomial
functions on $\C^2$ (ie in two variables) with the obvious $SU(2)$
action (or possibly the dual of the obvious one, depending on quite
how carefully you considered what ``obvious'' meant!)

Tensor products of such irreps are naturally spaces of holomorphic
sections of the external tensor product of these line bundles over a
product of Riemann spheres, for example
\[ V_a \otimes V_b \otimes V_c =
H^0(\PP^1 \times \PP^1 \times \PP^1, \LL^a \boxtimes \LL^b
\boxtimes \LL^c)\]
with the diagonal action of $SU(2)$ on spheres and bundles. 

The calculation we are going to perform is a stationary phase
integration, for which we need local differential-geometric
information about these holomorphic sections. We will take the
primarily symplectic point of view of Guillemin and Sternberg
\cite{GS}, as well as using the main theorem of their paper (see
below). Other relevant references are McDuff and Salamon \cite{MS} for
general symplectic background and for symplectic reduction, Kirillov
\cite{Kiri} for a concise explanation of geometric quantisation, and
Mumford et al \cite{MFK} for the wider context of geometric invariant
theory.

\subsection{K\"ahler geometry}

Suppose $M$ is a compact K\"ahler manifold of dimension $2n$. Thus, it
has a complex structure $J$ acting on the real tangent spaces $T_pM$,
a symplectic form $\omega$ and a Riemannian metric $B$ (we avoid the
symbol $g$, which will denote a group element). The latter are
$J$--invariant and compatible with each other according to the equation
\[ B(X,Y)= \omega (X, JY), \]
this being a positive-definite inner product. The Liouville volume
form $\Omega = \omega^n/n!$ equals the Riemannian volume form.
The hermitian metric on $T_pM$ (thought of as a complex space) is 
\[ H(X,Y)= B(X,Y) - i \omega(X,Y) \]
which is linear in the first factor and antilinear in the second (the
convention used throughout the paper).

\subsection{Hamiltonian group action}

Let $G$ be a compact group acting symplectically on $M$.  We assume
that the action is also Hamiltonian (ie that a moment map exists)
and that it preserves the K\"ahler structure. This will certainly be
true in the examples we will deal with.

We let $\lie{g}$ be the Lie algebra of $G$. Then an element $\xi \in
\lie{g}$ defines a Hamiltonian vector field $X_\xi$, a Hamiltonian
$\mu(\xi)$ and thus a moment map $\mu\co  M \ra \coo{g}$ according to the
conventions
\[ d\mu(\xi) = \iota_{X_\xi}\omega \quad (= \omega(X_\xi, -)). \]
I will later abuse notation slightly and write $\mu(X_\xi)$ instead of
$\mu(\xi)$ when I want to emphasise the association of the moment map
with a vector field corresponding to an infinitesimal action of $G$,
rather than with an explicit Lie algebra element.

\subsection{Equivariant hermitian holomorphic line bundle}

If the symplectic form $\omega$ represents an integral cohomology
class then there is a unique (we will assume $M$ is simply-connected)
hermitian holomorphic line bundle over $M$, with metric $\lan -,
-\ran$, whose associated compatible connection has curvature form
$F=(-2\pi i)\omega$ (so that $[\omega]$ is its first Chern class).

We assume that $G$ acts equivariantly on $\LL$, preserving its
hermitian form. The space of holomorphic sections $V=H^0(M, \LL)$ is
finite-dimensional and has a natural left $G$--action defined by 
\[ (gs)(p) = g.s(g^{-1}p).\]
$V$ becomes a unitary representation of $G$ when given the inner product
\[ ( s_1, s_2 ) = \int_M \lan s_1, s_2 \ran \Omega. \]
The round bracket notation will be used to distinguish the {\em
global} or {\em algebraic} hermitian forms from the {\em pointwise}
form on the line bundle $\LL$, which will be written with angle
brackets.

The infinitesimal action on sections is given by the formula
\begin{equation}\label{e:quant} \xi s = \frac{d}{dt}(\exp(\xi t)s(\exp(-\xi t)p))= (-\nabla_{X_\xi} +
2\pi i \mu(\xi) )s \end{equation} This is the fundamental
``quantization formula'' of Kostant et al.

\begin{rem} One has to be very careful
with signs here, especially as there is such variation of convention
in the literature. This formula is {\em minus} the Lie derivative
$\LL_{X_\xi}s$, because we are interested in the {\em left} action of
$G$, and the Lie derivative is defined using the contravariant (right)
action of $G$ on sections via pullback. There is an identical problem
if one looks at the derivative of the left action of $G$ on vector
fields, one has:
\[ \frac{d}{dt}(\exp(\xi t)_*Y)= - \LL_{X_\xi}Y= - [X_\xi,Y]\] 
provided one uses the standard conventions on Lie derivative and bracket: \[
[ X, Y] = \LL_XY, \qquad [X,Y]f = X(Yf)-Y(Xf) \] For further comments
on sign conventions see McDuff--Salamon \cite{MS}, remark 3.3, though
note that we do not here adopt their different Lie bracket
convention. Anyway, it is a good exercise to check that the formula
really does define a Lie algebra homomorphism: for this one also needs
the standard conventions on curvature:
\[ F(X,Y) = [ \nabla_X, \nabla_Y ] - \nabla_{[X,Y]} \]
and on Poisson bracket:
\[ \{f,g\}=- \omega(X_f, X_g)\]
\end{rem}

\subsection{Complexification}

In the Borel--Weil--Bott setup, one actually has a complex group $\GC$
acting on $\LL$ and $M$. (Of course it does not preserve the hermitian
structure on $\LL$, but its maximal compact part $G$ does.) The action
of $\clie{g}$ on sections of $\LL$ is given by
\begin{equation*} (i\eta)s = i (\eta s) = (-i \nabla_{X_\eta} -2\pi \mu(\eta)) s =
(-\nabla_{JX_\eta} -2\pi \mu(\eta)) s\end{equation*} the last identity coming
because $s$ is a holomorphic section, so is covariantly constant in
the antiholomorphic directions in $TM \otimes \C$:
\begin{equation*} \nabla_{X+iJX}s=0\end{equation*}
The point of this identity is that it gives us information about the
derivatives of an invariant section in directions orthogonal to the
slice $\mu^{-1}(0)$.

\subsection{Example}\label{exsph1} The $(k+1)$--dimensional irrep of $SU(2)$ is
obtained by quantising $S^2$ with a round metric and with an
equivariant hermitian line bundle $\LL^{\otimes k}$ of curvature
$k\omega$, where $\omega$ is the standard form with area 1. We will
always view $S^2$ as being the unit sphere in $\Rthree$. In
cylindrical coordinates, its unit area form is then (``Archimedes'
theorem'')
\[ \omega = \frac{1}{4\pi}d\theta \wedge dz. \]
Let $\xi$ be an element in the Lie algebra of the circle so that
$e^\xi=1$. The moment map for the $1$--periodic rotation about the
$z$--axis, generated by the vector field $X_\xi= 2\pi \partial/\partial
\theta$ is $\mu\co  S^2 \ra \R$ given by $\mu=\frac12 kz$. (Here we
identify the dual of the Lie algebra with $\R$ by letting $\xi$ be a
unit basis vector.)  Thus the image of the moment map is the interval
$[-\frac{k}{2},\frac{k}{2}]$, and in accordance with the
Duistermaat--Heckman theorem, the length of the interval equals the
area of the sphere.


\subsection{K\"ahler quotients}

Given as above a K\"ahler manifold $M$ and the K\"ahler, Hamiltonian
action of a compact group $G$, we may form the K\"ahler quotient
$M_G$, which is just an enhanced version of the symplectic
(Marsden--Weinstein) reduction.

Let $M_0$ denote the slice $\mu^{-1}(0)$. The $G$--equivariance of the
moment map $\mu\co  M \ra \coo{g}$ (coadjoint action on the right) means
that $M_0$ is $G$--invariant, and consequently that $\omega(X_\xi, Y)=
d\mu(\xi)(Y)=0$ for any $\xi \in \lie{g}$ and $Y \in T_{p}M_0$. Let us
suppose that $G$ acts freely on $M_0$, since this will be enough for
our purposes. We will use the symbol $\lie{g}p\equiv \{ X_\xi(p) : \xi
\in \lie{g} \}$ to denote the tangent space to the $G$--orbit at $p$
and similarly, $i\lie{g}p$ will denote $\{ JX_\xi(p) : \xi \in \lie{g}
\}$. In fact $\lie{g}p$ is the symplectic complement to $T_{p}M_0$ at
$p$, and $i\lie{g}p$ is the (Riemannian) orthogonal complement to
$T_{p}M_0$, because $B(Y, JX_\xi)=-\omega(Y, X_\xi)=0$ for any $Y \in
T_{p}M_0$, and the dimensions add up.

As a manifold, $M_G$ is just just the honest quotient $M_0/G$. It
inherits an induced symplectic form $\omega_G$ whose pullback to $M_0$
is the restriction of that of $M$. Its tangent space $T_{[p]}M_G$ at a
point $[p]$ (the orbit of a point $p\in M$) may be identified with its
natural horizontal lift, namely the orthogonal complement of
$\lie{g}{p} \subseteq T_{p}M_0$ at any lift $p$ of $[p]$. This space
may also be described as the orthogonal complement of $\clie{g}{p}
\subseteq T_{p}M$. As this subspace is complex, $T_{[p]}M_G$ inherits
both a Riemannian metric and a complex structure by
restriction. Hitchin proves in \cite{H} that these induced structures
make $M_G$ into a K\"ahler manifold. Starting from $\LL$ over $M$, we
can also construct a hermitian holomorphic line bundle $\LL_G$ over
$M_G$ with curvature $-2\pi i \omega_G$ (in particular, the induced
symplectic form is integral), as in \cite{GS}. The bundle and
connection are such that their pullback to $M_0$ agrees with the
restriction of $\LL$.

\subsection{Reduction commutes with quantization}

Let $Q(M)$ denote the {\em quantization} $H^0(M, \LL)$
associated to a K\"ahler manifold with equivariant hermitian line
bundle $\LL$ (which is suppressed in the notation). It is a
representation of $G$, so we can consider the space of invariants
$Q(M)^G \subseteq Q(M)$. (Whether we use $G$ or $\GC$ here is of
course irrelevant.)

The main theorem in \cite{GS} is that there is an isomorphism $Q(M)^G
\cong Q(M_G)$. There is obviously a restriction map from invariant
sections over $M$ to sections over $M_G$, more or less by definition
of $M_G$ and $\LL_G$, so the task is to show injectivity and
surjectivity. 

A vital ingredient in their proof of surjectivity is fact that norms
of invariant sections achieve their maxima (in fact decay
exponentially outside of) the slice $M_0$. We will rely on this fact
too. Suppose $s$ is a holomorphic $G$--invariant section over $M$, and
consider the real function $\norm s \norm^2$ on $M$. It is certainly
$G$--invariant, but not $\GC$--invariant. Following \cite{GS} we compute
the derivative
\begin{equation*}\label{e:norm} (JX_\eta)\norm s \norm^2 = -4 \pi \mu(\eta) \norm s
\norm^2 \end{equation*} by using the quantization formula
\eqref{e:quant} and the compatibility of hermitian metric and
connection
\[ X \norm s \norm^2 = \lan \nabla_X s, s \ran + \lan s,
\nabla_X s \ran. \] Therefore, if $\gamma(t)$ is the flowline
starting at $p \in M_0$ and generated by $JX_\eta$,
\[ \frac{d}{dt} \norm s \norm^2_{\gamma(t)} = -4\pi \mu(\eta) \norm s
\norm^2_{\gamma(t)} \]
and combining with 
\[ \frac{d}{dt} \mu(\eta)_{\gamma(t)}= B(X_\eta, X_\eta) > 0\] 
we see that indeed the function $\norm s \norm^2_{\gamma(t)}$ has a
single maximum at $t=0$, ie on $M_0$.

\subsection{Refinement of the Guillemin--Sternberg theorem}

We need an addition to the ``reduction commutes with quantization''
theorem. Any space $Q(M)$ is in a natural way a Hilbert space, with
inner product defined by
\[ (s_1,s_2) = \int_M \lan s_1, s_2 \ran \Omega\]
where $\lan - , - \ran$ is its line bundle's hermitian form. One might
imagine that the restriction isomorphism $Q(M)^G \cong Q(M_G)$ is an
isometry, but in fact it is not. However, asymptotically it becomes an
isometry if one redefines the measure on $M_G$, as will be shown
below.

First let us refine the observations about maxima of pointwise norms
of sections given above. We can repeat the argument using the
pointwise modulus of $\lan s_1, s_2 \ran $, and establish that its
maxima too are on $M_0$. Also, we can compute the second derivatives
of the function $\lan s_1, s_2 \ran $ in the $JX_\eta$ directions,
which span the orthogonal complement of $TM_0$. Redoing the above
calcuation yields, at $p \in M_0$,
\[ (JX_\xi.JX_\eta \lan s_1, s_2 \ran)_{p} = - 4 \pi JX_\xi(\mu(\eta)
\lan s_1, s_2 \ran)_{p}  = - 4 \pi B_{p}(X_\eta,
X_\xi)\lan s_1, s_2 \ran_{p}  \] because the moment map is zero at
$p$. Parametrising a regular neighbourhood of $M_0$ as $M \times
\exp(iU)$ for $U$ some small disc about the origin in $\lie{g}$, we
see that to second order the function satisfies
\begin{equation}\label{f:gauss} \lan s_1, s_2 \ran_{(p, \xi)} \approx \lan s_1, s_2
\ran_{p} e^{-2\pi
B(X_\xi, X_\xi)} \quad \hbox{for}\  \xi \in U.\end{equation}

To understand the asymptotics, we must first understand what is
varying!  Let $k$ be a natural number. Then one can consider $M$ with
the new symplectic form $k\omega$; its Liouville form scales by $k^n$
(recall $\dim M=2n$), the moment map for its $G$--action scales by $k$,
its Riemannian metric scales by $k$, and there is a new equivariant
hermitian holomorphic bundle $\LL^{\otimes k}$ over $M$ whose Chern
form is $k\omega$. If $s$ is a $G$--invariant section of $\LL$ then one
can consider $s^k = s^{\otimes k}$, which is an invariant section of
$\LL^{\otimes k}$.  We will always write the $k$ explicitly to
indicate the scaling of forms, so that $B$, $\omega$, $\Omega$,
$X_\xi$ and so on retain their original definitions. The new
pointwise hermitian form satisfies
\[ \lan s_1^k, s_2^k \ran = \lan s_1, s_2 \ran ^k.\]

Thus, in view of \eqref{f:gauss}, as $k \ra \infty$ the invariant
section has pointwise norm concentrating more and more (like a
Gaussian bump function) on the slice $M_0$. This localisation
principle rigorises Wigner's ideas and forms the basis for the proof
of the Ponzano--Regge formula.

\begin{theorem}\label{t:norm}
Let $\tilde s_1, \tilde s_2$ be $G$--invariant sections of $\LL$ over
$M$, and $s_1, s_2$ the induced sections of $\LL_G$ over $M_G$. Let
$\sigma\co M_G \ra \R$ be the function which assigns to a point $[p]$ the
Riemannian volume of the $G$--orbit in $M$ represented by $[p]$, and
let $d=\dim(G)$. Then, as $k \ra \infty$, there is an asymptotic
formula
\[(\tilde s_1^k, \tilde s_2^k) = \int_M \lan \tilde s_1^k, \tilde
s_2^k \ran (k^n\Omega) \sim \left(\frac{k}{2}\right)^{d/2} 
\int_{M_G} \lan s_1^k, s_2^k \ran (\sigma k^{n-d}\Omega_G). \]
\end{theorem}
\begin{proof}
From basic geometric invariant theory \cite{GS} we know 
$M_0/G=M_{ss}/\GC$, where $M_{ss}=\GC M_0$ is the set of semistable
points, an open dense subset of $M$. We view the left hand integral as
an integral over $M_{ss}$ and then integrate over the fibres of $\pi\co 
M_{ss} \ra M_G$, which are $\GC$--orbits. It helps to think of $\Omega$
as the Riemannian (coming from $B$) rather than the symplectic volume
form.

Suppose $[p]\in M_G$ and $p \in M_0$. The fibre can be parametrised
via the map $G \times i\lie{g} \ra \pi^{-1}([p])$ given by $(g, i\xi)
\mapsto \exp(i\eta)gp.$ This is a diffeomorphism because of the Cartan
decomposition of $\GC$. (Recall we are assuming that $G$ acts freely
on $M_0$.)

Let $\psi$ be the pullback to $G \times i\lie{g}$ of the function
$\log \lan \tilde s_1, \tilde s_2 \ran$, so that $\log \lan \tilde
s_1^k, \tilde s_2^k \ran$ pulls back to $k\psi$. It is invariant in
the $G$ directions but has Hessian form in the $i\lie{g}$ directions
(at $0$) given by
\[ (i\xi.i\eta.k\psi)_0 = -4\pi k B_{p}(X_\xi, X_\eta). \]
The pullback Riemannian metric on $i\lie{g}$ is given by $\beta(i\xi,
i\eta)= B(X_\xi, X_\eta)_{p}$. Consequently the integral of
$\e^{k\psi}$ over $i\lie{g}$ is asymptotically given by
\[ \lan \tilde s_1, \tilde s_2 \ran_{p}^k \int_{i\lie{g}}
e^{-2\pi k\beta(i\xi,i\xi)} d\vol_\beta = \lan \tilde s_1, \tilde s_2
\ran_{p}^k \left(\frac{\pi}{2\pi k}\right)^{d/2}.\] (The symbol
$\vol_\beta$ denotes the measure induced by the same metric $\beta$ as
appears in the integrand; changing coordinates to an orthonormal
basis, one obtains the a standard Gaussian integral, independent of
$\beta$.)

Finally we integrate over the $G$ orbit, picking up the factor
$\vol(Gp)=\sigma([p])$. Substituting into the original left-hand side
and separating the powers of $k$ in the correct way finishes the
proof.
\end{proof}

\subsection{Example}\label{exsph2} Let us return to example
\ref{exsph1} and check these formulae. If we have area form $2k\omega$
then sections of $\LL^{\otimes 2k}$ are identified as homogeneous
polynomials in $Z, W$ of degree $2k$, and under the circle action
there is an invariant section which one can write as
\[       s^k\co  (Z,W) \mapsto (Z^k W^k) \in \C. \]
In the complement of infinity, trivialise $\LL^{\otimes 2k}$ using the
nowhere-vanishing holomorphic section $b^{2k}$ corresponding to the
homogeneous polynomial $W^{2k}$. Let $\zeta=Z/W$ be the coordinate on
this chart. The pointwise norm of $b^{2k}$ is $(1+\vert \zeta
\vert^2)^{-k}$, because a unit element of the tautological bundle
above $\zeta$ is $(\zeta,1)/\sqrt{(1+\vert \zeta \vert^2)}$, which gets
sent to $(1+\vert \zeta \vert^2)^{-k}$ by the section $b^{2k}$.) Thus
the pointwise norm of $s^k$ is $\vert \zeta \vert^k/(1+\vert \zeta
\vert^2)^k$. Under stereographic projection
from the unit sphere in $\R^3$
\[ \zeta = \frac{x+iy}{1-z} \]
we get $1/(1+\vert \zeta \vert^2) = (1-z)/2$ and $\vert \zeta \vert
^2/(1+\vert \zeta \vert^2) = (1+z)/2$. Thus the pointwise norm-squared of $s^k$
is $((1-z^2)/4)^k$. Its global norm-square is 
\begin{eqnarray*}
\norm s^k \norm ^2 &= &\int_{S^2}  {\left(\frac{1-z^2}{4}\right)}^k
2k\frac{1}{4\pi}d\theta \wedge dz \\
&= & 2k \int_{-1}^1 {\left(\frac{1-z^2}{4}\right)}^k \frac12 dz\\
&=& 2k B(k+1, k+1)
\end{eqnarray*}
by definition of the beta function $B$. Evaluating the beta function
in terms of factorials gives
\[  \norm s^k \norm^2 =\frac{2k}{2k+1} {\binom{2k}{k}}^{-1} \sim
(\sqrt{\pi k})4^{-k}. \] Compare this with the computation of the norm
on the reduced space, which is a single point: one finds the
norm-square of $s^k$ at this point to be simply its value on the
equator $z=0$ of the sphere, namely $4^{-k}$, so the ratio of the two
is therefore $\sqrt{\pi k}$. If one applies theorem \ref{t:norm} with
$\Omega = 2 \omega$ then one gets the same ratio: the scaling factor
is $\sqrt{k/2}$ and the length of the equator (the factor $\sigma$ in
the formula) is $2\pi \sqrt{2/4\pi}$, because the usual spherical {\em
area} form has been divided by $4\pi$ and multiplied by $2$.

\subsection{Orbit volumes}
It is convenient here to note the formula for the volume of the
$G$--orbit with respect to some basis (we will eventually have to roll
up our sleeves and perform explicit calculations).

\begin{lemma}\label{t:orbvol}
If $\{\xi_i\}$ is a basis for and $\rho$ is an invariant metric on
$\lie{g}$, then the volume of the $G$--orbit at $p\in M_0$ is
\[ \vol(Gp) = \frac{\vol_{B_p}\{X_{\xi_i}\}}{\vol_{\rho}\{\xi_i\}}
\vol_{\rho}(G).\] 
\end{lemma}
\begin{proof}
Pulling back the metric via the diffeomorphism $G \ra Gp$, $g \mapsto
g{p}$ gives a metric on $G$ of the form
\[ \beta(g_*\xi, g_*\eta)=B_{g p}(X_\xi, X_\eta) = B_{p}(X_\xi, X_\eta)\] whose associated volume form differs from that of
$\rho$ by the given factor.
\end{proof}

It is also worth giving a formula for the orbit volume when $G$ does
not act freely on a space. Suppose as in the previous lemma that
$\rho$ is an invariant metric, that the stabiliser at $p$ is
$T$, whose Lie algebra is $\lie{t} \subseteq \lie{g}$. 

\begin{lemma}\label{t:rorbvol}
If $\{\xi_i\}$ is set of vectors in $\lie{g}$ which, when projected
onto into the orthogonal complement $\lie{t}^\perp \subseteq \lie{g}$,
forms a basis $\{\hat\xi_i\}$ for that space, 
\[ \vol(Gp) = \frac{\vol_{B_p}\{X_{\xi_i}\}}{\vol_\rho\{\hat\xi_i\}} \vol_{\rho}(G/T).\]
\end{lemma}
\begin{proof}
Repeat the earlier proof with $G/T$ mapping diffeomorphically to the
orbit, and using the basis $\{\hat\xi_i\}$ for the tangent space of
$G/T$. This gives a formula like the above except with
$\vol_{B_p}\{X_{\hat\xi_i}\}$ on top. However, since the vectors $\hat
\xi_i - \xi_i$ are in $\lie{t}$, they map to zero tangent vectors at
$p$, and we can simply remove the hats.
\end{proof}

\subsection{Stationary phase formulae}

The standard stationary phase formula is as follows: on a manifold
$M^{2n}$ with volume form $\Omega$, for a smooth real function $f$
with isolated critical points $\{p\}$, one has
\begin{equation*} \int_M e^{ikf}\Omega \sim \left(\frac{2\pi}{k}\right)^{n} \sum_{p}
\frac{e^{ikf(p)}.e^{\frac{i\pi}{4}
\sgn(\Hess_{p}(f))}}{\sqrt{\Hess_{p}(f)}}.\end{equation*}

In our computation of we will actually have a {\em complex} function
$\psi$, so it is probably easier to rewrite/generalise the above
formula to such a situation as
\begin{equation}\label{e:stat} \int_M e^{k\psi}\Omega \sim \left(\frac{2\pi}{k}\right)^{n} \sum_{p}
\frac{e^{k\psi(p)}}{\sqrt{-\Hess_{p}(\psi)}}\end{equation} where the
Hessian is now a complex number, and by the square root we mean the
principal branch (the Hessian must not be real and positive). Of
course, with $\psi = if$, $f$ real, this reduces to the previous
version.


\section{Warm-up example}

In order to demonstrate more clearly the main points of the
calculation to come in section 5, we will first work out a simpler
case. 

Let $V_{2k}$ be an irreducible representation of $SO(3)$, and let
$S^1_z$ be the circle subgroup fixing the $z$--axis in $\Rthree$. With
respect to this subgroup, $V_{2k}$ splits into one-dimensional weight
spaces indexed by even weights from $-2k$ to $2k$. We may pick unit
basis vectors inside these, uniquely up to a sign (by using the real
structure of the representation). If $g \in SO(3)$ is a rotation, we
may compute the matrix elements of $g$ with respect to such a basis by
using the hermitian pairing. Most of these depend on the choices of
sign, but the diagonal elements, those of the form $(v, gv)$, are
independent. Below we will compute an asymptotic formula for such a
matrix element, when $v$ is the zero-weight ($S^1$--invariant) vector.

There are in fact explicit formulae for matrix elements given using
Jacobi and Legendre polynomials, which are well-known in quantum
mechanics. One can prove the theorem from these more easily, see for
example Vilenkin and Klimyk \cite{VK}. Also, it could be computed
explicitly from the example \ref{exsph2}. But this method demonstrates
how to do the calculation without such explicit knowledge of the
sections.

If $v$ is a weight vector for $S^1_z$ then $gv$ is a weight vector for
$gS^1_zg^{-1} = S^1_{gz}$, the subgroup fixing the rotated axis
``$gz$''. The elements $(v,gw)$ can also be thought of as elements of
the matrix expressing one weight basis in terms of the other.

\begin{theorem}
As $k \ra \infty$ there is an asymptotic formula
\[ (v_0^{(k)}, gv_0^{(k)}) \sim \sqrt{\displaystyle\frac{2}{\pi k \sin \beta}}
\cos \left\{(2k+1)\frac{\beta}{2} + \frac{\pi}{4}\right\} \] where
$v_0^{(k)}$ is a unit zero-weight vector in $V_{2k}$ and $\beta>0$ the angle
through which $g$ rotates the $z$--axis.
\end{theorem}
\begin{proof}
As in example \ref{exsph1}, $V_{2k}$ is the space of holomorphic
sections of $\LL^{\otimes 2k}$ over $S^2$, with $SO(3)$ acting in the
obvious way, symplectic form $2k\omega$, and the subgroup $S^1_z$
acting with moment map $kz$. The zero-slice is a circle, and the
reduced space a single point. Therefore there is a one-dimensional
space of invariant sections, and such a section will have maximal (and
constant) modulus on the slice $z=0$.

We choose a section $s$ of $\LL^{\otimes 2}$ which is
$S^1_z$--invariant and has peak modulus $1$ where $z=0$. (The phase
doesn't matter, as noted above.) Then $s^{\otimes k}$ is a section of
$\LL^{\otimes 2k}$, invariant under $S^1_z$, and also with peak
modulus 1 at $z=0$. It does not quite represent a choice of $v_0^{(k)}$
because its global norm is not 1 (we have fixed it locally instead).

We must compute an asymptotic expression for 
\[ (v_0^{(k)}, gv_0^{(k)}) = \frac{(s^k, gs^k)}{(s^k,s^k)}.\]
The denominator is asymptotically $\sqrt{\pi k}$, as we checked in
example \ref{exsph2}. So the main work is computing the integral
\[ (s^k, gs^k) = \int_{S^2} \lan s^{k}, gs^{k} \ran
2k\omega = k\int_{S^2} \lan s, gs \ran^k (2\omega)=k \int_M
e^{k\psi}(2\omega)  \]
where $\psi = \log\lan s, gs \ran$.

Now $gs$ is an invariant section for $S^1_{gz}$, whose moment map is
simply the ``$gz$ coordinate'', and whose zero-slice ``$gz=0$'' meets
$z=0$ in two antipodal points $N$ and $S$. (They are both on the axis
of $g$, and $N$ is the one about which $g$ is anticlockwise rotation.)
Therefore, outside a neighbourhood of these two points, the modulus of
the integrand is exponentially decaying, and the asymptotic
contribution to the integral is just from $N$ and $S$.  In fact these
two points will also turn out to be the critical points of $\psi$, and
we will evaluate the integral using the standard stationary phase
procedure.

Let us denote by $\mu, \nu$ the moment maps for $S^1_z$ and $S^1_{gz}$
acting on the sphere with symplectic form $2\omega$. Let $X,Y$
be the generating vector fields corresponding to these actions, as
shown in figure \ref{sphere}.
\begin{figure}[ht]
\[ \vcenter{\hbox{\mbox{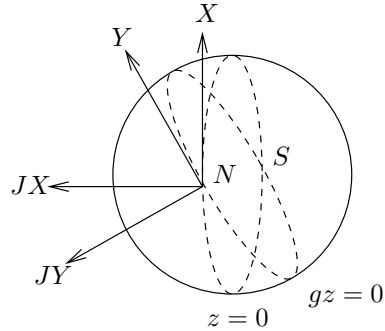}}}\]
\caption{Localisation regions and generating vectors\label{sphere}}
\end{figure}

The first derivatives of $\psi$ can be calculated as follows: start by
computing
\[ X \lan s, gs \ran =  \lan \nabla_{X}s, gs
\ran + \lan s, \nabla_{X} gs \ran.\] The first term can be simplified
to $2\pi i \mu \lan s, gs \ran$ via the quantization formula
\eqref{e:quant}, because $s$ is invariant under the group
corresponding to $X$. The second term is quite so easy, but becomes
simpler if we write \[X = p Y + q JY\] for some scalar functions $p,q$
(generically $Y, JY$ span the tangent space), and then expand again:
\[ X\psi = 2\pi i \mu -2\pi i p \nu - 2 \pi q\nu.\] 
At either intersection point, the moment maps are both zero, so the
whole thing vanishes. Similarly, the substitution
\[ Y = p' X - q' JX \]
gives
\[ Y \psi = 2\pi i p'\mu +2\pi q' \mu - 2 \pi i \nu.\]
Applying $X$ and $Y$ again to these formulae gives us the second
derivatives at the critical points. The fact that ultimately we
evaluate where the moment maps are zero shows we need only worry about
the terms arising from the Leibniz rule in which the vector field
differentiates them. These are evaluated using the following
evaluations at $N$:
\[ X\mu=0\qquad X\nu = 2\omega(Y,X)\qquad Y\mu = 2\omega(X,Y)\qquad Y\nu=0.\]
In addition, at $N$ we have $p=p'= \cos \beta$ and $q=q'=-\sin \beta$,
by inspection.  Thus, with respect to the basis $\{X, Y\}$, we have the
matrix of second derivatives
\[ (-2\pi i)(-2\omega(X,Y)) \left( \begin{matrix} e^{i\beta}&1\\
1&e^{i\beta}\end{matrix} \right). \] 
To obtain the Hessian, we have to divide the determinant of this
matrix by $(2\omega(X,Y))^2$, to account for the basis $\{X, Y\}$ not
being unimodular with respect to the volume form $2\omega$. Therefore
the Hessian at $N$ is
\[ (2\pi i)^2 e^{i\beta} 2i\sin \beta = -8\pi^2i\sin\beta
e^{i\beta}. \]
An easy check shows that the complex conjugate occurs at $S$. 

The $0$--order contributions $e^{k\psi(N)}, e^{k\psi(S)}$ must be
computed. Both are of unit modulus, because of our convention on
$s$. Their phases are easy to compute using the unitary equivariance
of the bundle. Let $h$ be the rotation through $\pi$ lying in
$S^1_z$. It exchanges $N$ and $S$, and preserves the section $s$. Thus
\begin{align*} e^{\psi(S)} & = \lan s,gs\ran(S) =  \lan h.h^{-1}s(hN), gh. h^{-1}s(hN) \ran = \lan
hs(N), ghs(N)\ran \\
&= \lan s(N), h^{-1}ghs(N)\ran.\end{align*} Now $h^{-1}gh$ is
clockwise rotation through $\beta$ at $N$, and therefore acts on the
fibre of $\LL^{\otimes 2}$, which is the tangent space at $N$, via
$e^{-i\beta}$. Hence
\[ e^{\psi(S)} = e^{i\beta}\ \text{and similarly}\ e^{\psi(N)}= e^{-i\beta} .\]

Finally we can put this all together:
\[ k\int_{S^2} e^{k\psi} (2\omega) = k \frac{2\pi}{k} \left\{
\frac{e^{-ik\beta}}{\sqrt{8\pi^2i\sin\beta e^{i\beta}}} +
\frac{e^{ik\beta}}{\sqrt{-8\pi^2i\sin\beta e^{-i\beta}}}\right\}.\]
Dividing by the normalising $(s,s)= \sqrt{\pi k}$ gives
\[ \sqrt{\displaystyle\frac{2}{\pi k \sin \beta}}
\cos \left\{(2k+1)\frac{\beta}{2} + \frac{\pi}{4}\right\}. \]
\end{proof}

\begin{rem}
The answer consists of a modulus, coming from the modulus of the
Hessian's determinant and the normalisation factors for the original
sections, and a phase, coming partially from the phase of the Hessian
and partially from the global phase shift (0--order terms). The $\pi/4$
is a standard stationary phase term. It will be possible to identify
the terms in the $6j$--symbol formula similarly.
\end{rem}

\begin{rem}
Other weight vectors behave similarly. If one repeats example
\ref{exsph2}, one finds beta integrals with more `$(1-z)$'s than
`$(1+z)$'s or vice versa, and the peak of the integrand also shifts to
the correct position: a circle of constant latitude whose constant $z$
coordinate equals the weight divided by $k$. Similar formulae for
general rotation matrix elements may be obtained.
\end{rem}


\section{Asymptotics of $6j$--symbols}

\subsection{Geometry of the sphere}\label{51}

The $6j$--symbol arises by pairing two $SU(2)$--invariant vectors in a
12--fold tensor product of irreducibles, according to definition
\ref{6jdef}. Let us first consider the geometry of single
irreducibles.

Identify the Lie algebra of $SO(3)$ with $\Rthree$ by using the
standard vector product structure ``$\times$'', so that the standard
basis vectors $e_i$ generate infinitesimal rotations $v \ra e_i \times
v$ in space. Define an invariant metric $\rho$ (the usual scalar
product ``$.$'') by making them orthonormal. Using this scalar product
we identify $\Rthree$ with the coadjoint space also.  This
metric gives the flows generated by the `$e_i$'s period $2\pi$, and so
gives the circle $T$ in $SO(3)$ generated by any of them length
$2\pi$. We can think of $SO(3)$ as half of a 3--sphere whose great
circle has length $4\pi$; such a sphere has radius $r=2$, and
therefore volume $2\pi^2r^3 = 16 \pi^2$. Therefore $SO(3)$ has volume
$8\pi^2$, and the quotient sphere $SO(3)/T$ (with its induced metric) has
volume $4 \pi$.

The irrep $V_a$ is the space of holomorphic sections of the bundle
$\LL_a$ on $S^2$. To write down explicit formulae for the various
structures on the sphere it is better to view it as the sphere $S^2_a$
of radius $a$ in $\Rthree$, instead of as the unit sphere. The
symplectic form with area $a$ is given by
\begin{equation*} \omega_x(v,w) = \frac{1}{4\pi a^2} [x.v.w]\end{equation*}
where $x$ is a vector on the sphere, $v,w$ are tangent vectors at $x$
(orthogonal to it as vectors in $\Rthree$) and the square brackets
denote the triple product
\[ [x.v.w] \equiv x . (v \times w). \]
The complex structure at $x$ is the standard rotation
\begin{equation*} J_x(v) = \frac{1}{a} x \times v .\end{equation*}
The final piece of K\"ahler structure, the Riemannian metric, is then 
\begin{align*} B_x(v,w)& = \omega_x(v,J_xw) = \frac{1}{4\pi a^3} (x \times v).(x
\times w) = \frac{1}{4\pi a^3} (x^2 (v.w) - (x.v)(x.w))\\ & =
\frac{1}{4\pi a} v.w.\end{align*} This formula agrees with the fact
that with the {\em natural} induced metric given by $B_x(v,w)=v.w$,
the area of $S^2_a$ is $4 \pi a^2$.

The group $G=SO(3)$ acts on the sphere, preserving its K\"ahler
structure. This is a Hamiltonian action with moment map being simply
inclusion $\mu_a\co  S^2_a \hra \R^3$. If $a$ is even then $G$ also acts
on the corresponding hermitian line bundle, but if $a$ is odd one gets
an action of the double cover $SU(2)$. (Since we need to work primarily
with the geometric action, it is $SO(3)$ that wins the coveted title
``$G$''!)

Products of K\"ahler manifolds have sum-of-pullback symplectic forms,
and direct sums of complex structures. Their Liouville volume forms
are wedge-products of pullbacks (there are no normalising factorials;
this is one reason for the $n!$ in the definition of the Liouville
form). The moment map for the diagonal action of $G$ on $S^2_a \times
S^2_b \times S^2_c$ is therefore $\mu(x_1, x_2, x_3)=x_1+x_2+x_3 \in
\Rthree$. We know from the discussion earlier that the pointwise norm
of an invariant section over this space will attain its maximum on the
set $\mu=0$, which is in this case the ``locus of triangles''
$\{x_1+x_2+x_3=0\}$. $G$ acts freely and transitively on this space,
except in the exceptional cases when one of $a,b,c$ is the sum of the
other two. We can safely ignore this case, because it corresponds to a
flat tetrahedron (about which the main theorem says nothing).

We need to pick a section $s^{abc}$ corresponding to the invariant
vector $\epsilon^{abc}$, whose normalisation was explained in section
2. There is a one-dimensional vector space of $SU(2)$--invariant
sections of $\LL^{\otimes a} \boxtimes \LL^{\otimes b} \boxtimes
\LL^{\otimes c}$ over $S^2_a \times S^2_b \times S^2_c$. First we
define $s^{abc}$ uniquely up to phase by setting its peak modulus to
be 1. This section does {\em not} represent $\epsilon^{abc}$ exactly,
because we fixed the norm {\em locally}, instead of globally. However,
it is more convenient for calculations, and we can renormalise
afterwards. Similarly, let $s^{aa}$ be a section over $S^2_a \times
S^2_a$ with peak modulus 1. It will be concentrated near the zeroes of
the moment map $\mu(x_1, x_2)=x_1+x_2$, namely the anti-diagonal. Note
that $SO(3)$ acts transitively on the antidiagonal with circle
stabilisers everywhere.  To represent $\epsilon^{aa}$, this section
would have to be renormalised so that its global section norm was
$\sqrt{a+1}$.

Fixing the phase of each section is slightly more subtle. The phase of
the `$\epsilon$'s was fixed by using the spin-network
normalisation. The analogue for sections is just the same, viewing the
Riemann sphere as $\PP^1$ with homogeneous coordinates $Z$ and $W$,
and defining the sections just as before. The choice will not actually
matter until subsection \ref{phase}.


\subsection{A 24--dimensional manifold}

Fix 6 natural numbers $a,b, \ldots f$ satisfying the appropriate
admissibility conditions for existence of the $6j$--symbol.

We will work on the following K\"ahler manifold $M$:
\[ M= S^2_a \times S^2_b \times S^2_c \times S^2_c \times S^2_d \times
S^2_e \times S^2_e \times S^2_f \times S^2_a \times S^2_f \times S^2_d
\times S^2_b\] 
This is taken to lie inside $(\Rthree)^{12}$, and a point in it will
be written as a vector $(x_1, x_2, \ldots,  x_{12})$.

There are three useful actions on $M$. First there is the diagonal
action of $G$ on all 12 spheres. It has moment map $\phi\co  M \ra
\Rthree$ given by
\[ \phi(x_1, x_2, \ldots, x_{12}) =\sum_1^{12} x_i. \]
Algebraically, this generates the diagonal action of $G$ on the
corresponding tensor product of 12 irreducible representations
\[ V_a \otimes V_b \otimes V_c \otimes V_c \otimes \cdots \otimes V_b.\]

Secondly, one has an action of $G^4=G \times G \times G \times G$, the
first copy acting diagonally on the first three spheres, the next on
the next three, and so on. The moment map for this action is $\mu\co  M
\ra (\Rthree)^4$, \[ (x_1, x_2, \ldots, x_{12})\mapsto (x_1+x_2+x_3,
x_4+x_5+x_6, x_7+x_8+x_9, x_{10}+x_{11}+x_{12}).\] The section $\tilde
s_\mu= s^{abc} \boxtimes s^{cde} \boxtimes s^{efa} \boxtimes
s^{fdb}$ is a well-defined (given earlier conventions on $s^{abc}$)
invariant section for this action with peak modulus 1.

Thirdly, we let $G^6$ act on $M$, each copy acting diagonally on a
pair of spheres of the same radius. The moment map (which shows
precisely how this works) is $\nu\co  M \ra (\Rthree)^6$,
\[ \nu(x_1 , x_2, \ldots, x_{12})= (x_1+x_9, x_2+x_{12}, x_3+x_4, x_5+x_{11},
x_6+x_7, x_8+x_{10}).\] Then $\tilde s_\nu= s^{aa} \boxtimes s^{bb}
\boxtimes \cdots \boxtimes s^{ff}$ (after a suitable permutation
of its tensor factors, so that $s^{aa}$ lives over the first and ninth
spheres, for example) is an invariant section for this action with
peak modulus 1.

\subsection{Proof of theorem \ref{t:main}}

Recall from \ref{6jdef} the definition of the $6j$--symbol as a
hermitian pairing of two vectors, and their normalisations. The
corresponding geometric formula, in terms of the sections $\tilde
s_\mu, \tilde s_\nu$ just defined is
\begin{equation}\label{e:main} \ksixj = (-1)^{\sum a}\frac{(\tilde s_\mu^k,
\tilde s_\nu^k)}{\norm \tilde s_\mu^k \norm \norm \tilde
s_\nu^k\norm}\left(\prod (ka+1)\right) \end{equation} where the
product on the right denotes simply $(ka+1)(kb+1) \cdots (kf+1)$.) Our
convention on peak modulus 1 and phase of the sections mean that
$(s^{abc})^{\otimes k}=s^{ka,kb,kc}$ and similarly $(s^{aa})^{\otimes
k}=s^{ka,ka}$, so this formula is just the pairing of tensor products
of these sections, with corrections for the global norms of $\tilde
s_\mu$ and $\tilde s_\nu$ as explained at the end of subsection
\ref{51}.

To extract the asymptotic formula for the $6j$--symbol, we therefore
need asymptotic formulae as $k \ra \infty$ for the three integrals:
\begin{eqnarray*} I &= &(\tilde s_\mu^k, \tilde s_\nu^k)
= \int_M \lan \tilde s_\mu^k, \tilde s_\nu^k \ran k^{12}\Omega\\
I_\mu &=& (\tilde s_\mu^k, \tilde s_\mu^k) = \int_M \lan \tilde
s_\mu^k, \tilde s_\mu^k \ran k^{12}\Omega\\ I_\mu &= &(\tilde
s_\nu^k, \tilde s_\nu^k) = \int_M \lan \tilde s_\nu^k, \tilde s_\nu^k
\ran k^{12}\Omega  \end{eqnarray*} (Note the explicit inclusion of
all factors of $k$; everything else is unscaled.) We can in fact
immediately write down asymptotic formulae for the correction
integrals $I_\mu, I_\nu$ using theorem \ref{t:norm}, because the
reductions $M_{G^4}$, $M_{G^6}$ are both single point spaces, and the
sections have modulus $1$ over these points.
\begin{eqnarray}\label{e:gnorm}
I_\mu &\sim& \left(\frac{k}{2}\right)^6 \vol(\mu^{-1}(0))\\
\label{e:gnorm2}I_\nu &\sim& \left(\frac{k}{2}\right)^6 \vol(\nu^{-1}(0))
\end{eqnarray}
(In the second case one must actually reconsider the proof of the
theorem, because $G^6$ does not act freely on the set $\nu^{-1}(0)$,
but there is no problem.)

The remaining integral $I$ is evaluated by reduction to an integral
over $M_G$ followed by the method of stationary phase, which fills the
rest of this section.

\subsection{Localisation of the integral $I$}

As $k \ra \infty$, the integrand decays exponentially outside the
region where both moment maps $\mu, \nu$ are zero, because it is
dominated by the pointwise norms of the invariant sections $\tilde
s_\mu, \tilde s_\nu$. What then is the set $\mu^{-1}(0) \cap
\nu^{-1}(0)$?

\begin{figure}[ht]
\[ \vcenter{\hbox{\mbox{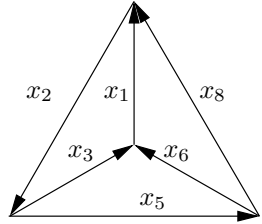}}}\]
\caption{Schematic configuration of vectors\label{schem}}
\end{figure}

At a point $(x_1, x_2, \ldots, x_{12})$, the condition $\nu=0$
requires that six of the $x_i$'s are simply negatives of the other
six, and then $\mu=0$ forces the six remaining ones, say $(x_1, x_2,
x_3, x_5, x_6, x_8)$, to form a tetrahedron, shown {\em schematically}
in figure \ref{schem}. Recall that the lengths of the vectors are fixed integers
$a,b,c,d,e,f$.

We have assumed that the numbers $a,b, \ldots, f$ satisfy the triangle
inequalities in triples (otherwise the $6j$--symbol is simply zero), so
the faces of this triangle can exist {\em individually} in
$\Rthree$. However, it is still quite possible that there is no
Euclidean tetrahedron $\tau$ with sides $a,b, \ldots, f$. The sign of
the Cayley polynomial $V^2(a^2, b^2, \ldots, f^2)$ (whose explicit
form is irrelevant here) is the remaining piece of information needed
to determine whether $\tau$ is Euclidean, flat or Minkowskian. In the
last case, we see that $\mu^{-1}(0) \cap \nu^{-1}(0)=\emptyset$, and
so have proved the second part of the main theorem: that if $\tau$ is
Minkowskian then the $6j$--symbol is exponentially decaying as
$k\ra\infty$.

\begin{figure}[h]
\[ \vcenter{\hbox{\mbox{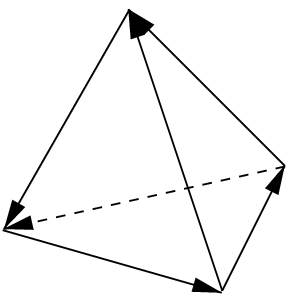}}} \qquad
\vcenter{\hbox{\mbox{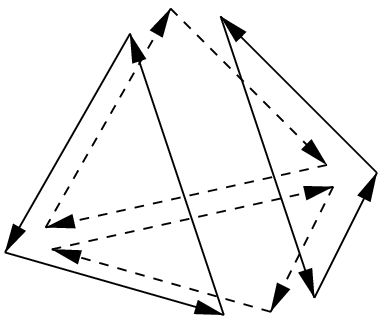}}}\]
\caption{Actual configuration of vectors\label{f:vect}}
\end{figure}

Suppose on the other hand that $V^2$ is positive. Then we {\em can}
find a set of six vectors $a,b,c,d,e,f$ in $\Rthree$ forming a
tetrahedron oriented as shown in figure \ref{f:vect}. Let $\tau$
denote both this tetrahedron and the corresponding point
\[ (a,b,c,-c,d,e,-e,f,-a,-f,-d,-b)
\in M \] and $\tau'$ be its mirror image (negate these 12
vectors). (Of course the whole tetrahedron is determined by just three
of the vectors, say $a,c,e$.)  It is clear that the localisation set
$\mu^{-1}(0) \cap \nu^{-1}(0)$ will consist of exactly two $G$--orbits
$G\tau$, $G\tau'$.

\begin{rem}
The symbols $a,b,c,d,e,f$ now denote {\em both} vectors and their
integer lengths at the same time! This ought not be {\em too}
confusing, as it should be clear from formulae what each symbol
represents. 
\end{rem}

Returning to the integral $I$, since both $\tilde s_\mu^k, \tilde
s_\nu^k$ are invariant under the diagonal action of $G$, we can apply
theorem \ref{t:norm} with respect to the diagonal action of $G$, and
obtain an integral over an 18--dimensional manifold $M_G$:
\begin{equation} \label{e:red} I = \int_M \lan \tilde s_\mu^k, \tilde s_\nu^k\ran k^{12} \Omega =
k^9\left(\frac{k}{2}\right)^{3/2} \int_{M_G} \lan s_\mu^k, s_\nu^k
\ran \sigma \Omega_G\end{equation} where in the right-hand integral,
$s_\mu, s_\nu$ are the descendents of $\tilde s_\mu, \tilde s_\nu$,
and the only thing depending on $k$ is the integrand, which is the
$k$th power of something independent of $k$. The function $\sigma$ is
the function on $M_G$ giving the volume of the corresponding $G$--orbit
in $M$, and we view it as part of the measure in the integral.

Let us define $\psi = \log \lan s_\mu, s_\nu \ran$ on $M_G$ and
$\tilde \psi = \log \lan \tilde s_\mu, \tilde s_\nu \ran$ on $M$, so
that the remaining problem is to compute 
\[ I' = \int_{M_G} e^{k\psi} \sigma \Omega_G.\]
Since the modulus of $e^{k\tilde\psi}$ localises on the set
$\mu^{-1}(0) \cap \nu^{-1}(0)= G\tau \cup G\tau' \subseteq M$, the
above integrand localises to the two {\em points} $[\tau], [\tau']$.

\subsection{Tangent spaces and stationary phase}

The diagonal $G$--action does not commute with the other
ones, so that $M_G$ will not have any kind of induced actions of $G^4$
or $G^6$, but we don't need this for the localisation calculation to
go through. We always work on the upstairs space $M$ not $M_G$,
precisely because the presence of the group actions defining the
invariant sections being paired is so useful.
The tangent space $T_{\tau}M$ is 24--dimensional, and contains
two 12--dimensional subspaces $\ker d\mu$ and $\ker d\nu$ which meet in
the 3--dimensional space $\lie{g}\tau$. (This degree of
transversality can be checked explicitly from formulae below, but it
should be clear from the fact that there are just two isolated
critical points in $M_G$.) Together they span the 21--dimensional
$T_{\tau}M_0$, which is orthogonal to $i\lie{g}\tau$.

Projecting to 18--dimensional $T_{[\tau]} M_G$, we see two 9--dimensional
subspaces we shall call $W_\mu$ and $W_\nu$ (the projections of $\ker
d\mu$ and $\ker d\nu$) meeting transversely at the origin.

We want to examine the behaviour of $\psi$ (its gradient and Hessian)
at the point $[\tau] \in M_G$. Let us choose orthonormal bases $\{X_1,
X_2, \ldots, X_9\}, \{Y_1, Y_2, \ldots, Y_9\}$ for the transverse
9--dimensional tangent spaces $W_\mu, W_\nu$ inside
$T_{[\tau]}M_G$. Then we can need to compute quantities such as $X_i
\psi$ and $X_i Y_j \psi$ (also at $[\tau]$, of course!). These can be
computed by choosing {\em arbitrary} lifts of the vectors to
$T_{\tau}M$ and applying them to the $G$--invariant function
$\tilde \psi =\log \lan \tilde s_\mu, \tilde s_\nu \ran$ on $M$.  This
is important, because it is very hard to write down any explicit {\em
horizontal} lifts which would be needed to do computations directly
in $T_{[\tau]}M_G$.

So, to compute something like $X_i \psi$ one can choose any lift
$\tilde X_i$ inside $\ker d\mu$ in $T_{\tau}M$, and write
\begin{eqnarray*}
X_i \psi &=& \tilde X_i \tilde \psi = \tilde X_i \log \lan
\tilde s_\mu, \tilde s_\nu \ran\\
&= & \lan
\tilde s_\mu, \tilde s_\nu \ran^{-1}( \lan
\nabla_{\tilde X_i} \tilde s_\mu, \tilde s_\nu \ran +
\lan \tilde s_\mu, \nabla_{\tilde X_i} \tilde s_\nu \ran). 
\end{eqnarray*}
Now $\tilde X_i$ is a generator of the $G^4$ action under which
$\tilde s_\mu$ is invariant, and therefore
\[ \nabla_{\tilde X_i}\tilde s_\mu = 2\pi i \mu(\tilde X_i) \tilde
s_\mu \] which vanishes at $\tau$.  (Here $\mu(\tilde X_i)$ really
denotes $\mu(\xi_i)$ for the Lie algebra element $\xi_i$ corresponding
to $\tilde X_i$.)  For the second term we must first express $X_i$ as
a linear combination of the $Y_j$ and $JY_j$ (which span
$T_{[\tau]}M_G$), then we can lift and use the quantization formula to
compute.

Therefore, introduce the $9 \times 9$ matrices $P_{ij}$ and $Q_{ij}$
according to 
\begin{align*}
X_i& = \sum P_{ik} Y_k + \sum Q_{ik} JY_k.\\
\intertext{Multiplying by $J$ we get}
JX_i &= - \sum Q_{ik} Y_k + \sum P_{ik} JY_k .
\end{align*}
By applying $\omega_G(X_j, -)$ and similar operators to these equations
one obtains
\[ P_{ij}= B_G(X_i, Y_j) \qquad Q_{ij}=-\omega_G(X_i, Y_j).\]
These, together with the fact that the bases are orthonormal and span
isotropic subspaces, determine completely matrices for $B$ and
$\omega$ on $T_{[\tau]}(M_G)$.  By a similar procedure one can invert
the relations:
\begin{eqnarray*}
Y_i &= &\sum P_{ki} X_k - \sum Q_{ki} JX_k \\
JY_i& = & \sum Q_{ki} X_k + \sum P_{ki} JX_k.
\end{eqnarray*}
A final point is that since $\{X_i, JX_i\}$ and $\{Y_i, JY_i\}$ are
both complex-oriented orthonormal bases for $T_{[\tau]}(M_G)$, the
change of basis matrix is special orthogonal, and hence
\[ P^TP+Q^TQ=1 \qquad QP^T=PQ^T \qquad Q^TP=P^TQ.\]

Now we may rewrite the tangent vectors appropriately, lift everything
to $T_{\tau}M$ and then apply them to $\tilde \psi$ via the fundamental
formula (recall $\lan - , -\ran$ is conjugate linear in the second
factor). For example
\begin{align*}
\tilde X_i \tilde \psi &= 2\pi i \mu(\tilde X_i) -2\pi i
\sum P_{ik} \nu(\tilde Y_k) -2\pi \sum Q_{ik} \nu(\tilde Y_k).\\
\intertext{This right hand side vanishes at $\tilde\tau$, so indeed $ X_i \psi
=0$ there.  The companion formula is}
\tilde Y_i \tilde \psi &= 2\pi i \sum P_{ki} \mu(\tilde X_k) +2\pi
\sum Q_{ki} \mu(\tilde X_k) -2\pi i \nu(\tilde Y_i).
\end{align*}
Together, these show that $\psi$ is stationary at $[\tau] \in M_G$,
just as in the warm-up example.

\subsection{Computation of the Hessian}
Another application of the above formulae, remembering that \[ X\mu(Y)=
d\mu(Y)(X)=\omega(Y,X)\] will obtain formulae for second derivatives
such as
\[ \tilde X_j \tilde X_i \tilde \psi = 2\pi i \omega_G(X_i, X_j) -2\pi i \sum
P_{ik} \omega_G (Y_k, X_j) -2\pi \sum Q_{ik} \omega_G(Y_k, X_j)\]
where everything in this formula is to be evaluated at $\tau$ (for
example, the first term now dies), and we have used the defining
identity $\omega(\tilde X, \tilde Y) = \omega_G(X, Y)$ to replace
$\omega$ by $\omega_G$ and remove the tildes from the right-hand side.

We can compute three similar formulae for the second derivatives and
form the Hessian matrix for $\psi$ with respect to the basis $\{X_i,
Y_i\}$ of $T_{[\tau]}M_G$ :
\[ (-2 \pi i) \left( \begin{matrix} PQ^T-iQQ^T&Q\\ Q^T&P^TQ-iQ^TQ
\end{matrix} \right) \]
We can extract the matrix
\[\left( \begin{matrix} Q^T&0\\ 0&Q
\end{matrix} \right)\] form the right, and expand  
\[ \det \left( \begin{matrix} P-iQ&1\\ 1&P^T-iQ^T
\end{matrix} \right)\]
as 
\[ \det( (P-iQ)(P^T-iQ^T)-1) = \det (-2QQ^T-2iPQ^T) \]
using properties of $P$ and $Q$ discussed earlier.  Hence this
temporary ``unnormalised Hessian'' of $\psi$ is:
\[ (-2 \pi i)^{18}. (-2i)^9 .\det(P-iQ). \det(Q)^3\]
The reason for separating the last two parts is that $\det Q$ is real,
whereas $P-iQ$ represents the change of basis between $\{X_i\}$ and
$\{Y_j\}$ as bases of $T_{[\tau]}M_G$ as a 9--dimensional {\em complex}
vector space (ie, it is the matrix of the hermitian form,
$(P-iQ)_{ij}= \overline{H_G(X_i, Y_j)}$), so is unitary and
contributes just a phase as determinant.

To normalise the Hessian we must compute the volume of the basis
$\{X_i, Y_j\}$ with respect to the form $\sigma\omega_G^9/9!$ on
$T_{[\tau]} M_G$. Expand, using the shuffle product, the expression
$(\omega_G^9/9!)(X_1, X_2, \ldots, X_9, Y_1, Y_2, \ldots, Y_9)$: since
the spaces spanned by the $X_i$ and by the $Y_j$ are isotropic, the
terms appearing are simply all possible orderings of all possible
products of 9 terms of the form $\omega(X_i, Y_j)$ (the $X$ before the
$Y$). Reordering these cancels the denominator $9!$ and we obtain
simply $\det(Q)$. So the determinant we actually computed was
$(\vol(G\tau)\det(Q))^2$ times what it should have been when computed
in a unimodular basis. Therefore
\[ \Hess_{[\tau]}(\psi)=  (-2 \pi i)^{18}. (-2i)^9. \det(P-iQ)
.\det(Q).\vol(G\tau)^{-2}.\]

This is the end of the general nonsense. To go any further we have to
choose explicit bases, although not for $T_{[\tau]} M_G$, because of
the difficulties already mentioned in writing down {\em any} vectors
there. In the next two sections we will write down nice vectors
``upstairs'' in $T_\tau M$ and show how to lift the computations of
$\det(P-iQ), \det(Q)$ into this space.

\subsection{The modulus of the Hessian}

We need to compute $\det(Q)$, where $Q_{ij}=-\omega_G(X_i, Y_j)$, and
the $X_i$, $Y_j$ are orthonormal bases as chosen above. 

Let us start by introducing some useful vectors in $T_\tau M$, with
which to compute ``upstairs''. We make an explicit choice of basis for
each of the 12--dimensional spaces $\ker d\mu, \ker d\nu$ inside
$T_{\tau}M$. Recalling that they intersect in the space $\lie{g}\tau$,
we arrange for a suitable basis of this space to be easily obtained
from each.

Let $T^l_v$ be the infinitesimal rotation about the vector $v$, acting
on the $l$th triple of vectors from $(x_1, x_2, \ldots, x_{12})$. For
example, at any point $(x_1, x_2, \ldots, x_{12})$, we have
\[ T^1_v=(v \times x_1, v \times x_2, v \times x_3, 0,0,0,0,0,0,0,0,0)
.\] This vector clearly preserves the condition $x_1+x_2+x_3=0$, as
well as the other three $\Rthree$--coordinate parts of $\mu$. 

Recall that $a,c,e$ are three vectors defining the tetrahedron $
\tau$. Since $a,c,e$ are linearly independent, the vectors $T^1_a,
T^1_c, T^1_e$ span the tangent space to $\{x_1+x_2+x_3=0\}$ inside the
product of the first three spheres of $M$. Combining four such sets of
vectors, we see that the 12 vectors $T^l_v$, where $l=1,2,3,4$ and $v$
is one of the three vectors $a,c$ or $e$, span the space $\ker d\mu$
at $\tau$. For convenience these vectors will also be numbered
\[T_1, T_2, \ldots, T_{12} = T^1_a, T^1_c, T^1_e, T^2_a, \ldots,
T^4_e.\] Note that although the formula defines a vector field
everywhere on $M$, we only need the tangent vectors at two specific
points, namely $\tau$ and $\tau'$.

We can easily obtain a basis for the infinitesimal diagonal action of
$G$ from these: \[R_a=T^1_a+T^2_a+T^3_a+T^4_a\] is the infinitesimal
rotation of all 12 coordinates about $a$, and similarly we may define
$R_c, R_e$, each a sum of four `$T$'s, which together span
$\lie{g}\tau$. We will also denote these by 
\[ R_1, R_2, R_3 = R_a, R_c, R_e. \]

Let $u$ denote an edge of the tetrahedron $\tau$, one of the vectors
$a,b,c,d,e,f$. Let $U^u_w$ be the infinitesimal rotation about $w$
acting on the pair of spheres corresponding to $u$. For example, if
$u=a$ then we have at $(x_1, x_2, \ldots, x_{12})$
\[ U^a_w=(w \times x_1, 0,0,0,0,0,0,0,w \times (-x_1),0,0,0)
,\] This vector preserves $x_1+x_9=0$ and hence $\nu$, and so do the
other $U^u_w$. We want just two vectors $w_1, w_2$ such that
$U^a_{w_1}, U^a_{w_2}$ span the tangent space to the orbit of $G$
acting on the first and ninth spheres in $M$ at $\tau$ (compare
the previous case with the three `$T$'s.) Projecting into the
first and ninth spheres, $\tau$ becomes $(a, -a)$ and $U^a_w$
becomes the tangent vector $(w \times a, w \times (-a))$. So all we
need to do is pick $w_1, w_2$ such that $a, w_1, w_2$ are linearly
independent. In this way we can construct 12 vectors spanning $\ker
d\nu$ at $\tau$. 

Unfortunately there isn't a totally systematic way of deciding which
two values of $w$ we should use, given $u$. We can at least choose
them always to be two of the three vectors $a,c,e$, which forces for
example the use of $U^a_c, U^a_e$ among our 12 vectors (because
$U^a_a=0$).

The twelve explicit choices are as follows:
\[ U_1, U_2, \ldots, U_{12} = U^a_c, U^a_e,  \quad U^b_a, U^b_e, \quad U^c_e, U^c_a, \quad U^d_c, U^d_a, \quad U^e_a, U^e_c, \quad U^f_c, U^f_e \]
The same three diagonal generators $R_a, R_c, R_e$ can be expressed in
terms of these vectors by observing that
\[ R_a = U^a_a+U^b_a+U^c_a+U^d_a+U^e_a+U^f_a \]
that $U^a_a=0$ and that $U^f_a$ (which is the only other not among our
chosen $U_1, U_2, \ldots, U_{12}$) satisfies $U^f_a=-U^f_e$, because
the fact that the three sides of the tetrahedron $\tau$ satisfy
$-e+f-a=0$ implies
\[ U^f_e+U^f_a=U^f_e-U^f_f+U^f_a=U^f_{e-f+a}= U^f_0 = 0.\]
Similarly, one obtains $U^b_c=-U^b_a$ and $U^d_e=+U^d_c$,
and hence:
\begin{eqnarray*}
R_a &=& U_3+U_6+U_8+U_9-U_{12} \\
R_c &=& U_1-U_3+U_7+U_{10}+U_{11} \\
R_e &= &U_2+U_4+U_5+U_7+U_{12}
\end{eqnarray*}

In the following calculation, a symbol such as
$\det\omega(\{X_i\};\{Y_i\})$, where $\{X_i\}$ and $\{Y_i\}$ are some sets
of vectors, will mean the determinant of the matrix whose entries are
all evaluations of $\omega$ on pairs consisting of an element from the
first set followed by one from the second set (arranged in the obvious
way). In the case where the two sets of vectors are both bases of some
fixed vector space, the symbol $\det(\{X_i\}/\{Y_i\})$ will be the
determinant of the linear map taking $Y_i \mapsto X_i$. The
grossly-abused subscript $i$ below stands for the complete list of
such vectors (there are twelve `$T$'s, three `$R$'s, etc.) We regard
all vectors as living in $T_\tau M$, in particular the original
orthonormal bases are lifted horizontally into it. Let $\{e_1, e_2,
e_3\}$ be some orthonormal basis of $\lie{g}\tau$.

By extending the orthonormal sets of vectors and then changing bases
inside the spaces $\ker d\mu$ and $\ker d\nu$ to bring in the `$T$'s and
`$U$'s , we have
\begin{align*}
\det(Q) &=-\det\omega(\{X_i\};\{Y_i\})\\
&= -\det\omega(\{X_i, e_i, Je_i\};\{Y_i, e_i, Je_i\})\\
&= -\det(\{T_i\}/\{X_i, e_i\})^{-1}\det(\{U_i\}/\{Y_i,
e_i\})^{-1}\\
&\qquad\times \det\omega(\{T_i, Je_i\};\{U_i, Je_i\})
\end{align*}
The remaining $\det \omega$ term can be simplified further. Replace
three of the `$T$'s ($T_{10}, T_{11}, T_{12}$) and three `$U$'s ($U_8,
U_{11}, U_4$) by $R_1, R_2, R_3$. According to the earlier expressions
for the $R_i$, each replacement is unimodular, and there is no sign
picked up in reordering the `$U$'s to put $U_8, U_{11}, U_4$ (in that
order) last. Then we change the `$R$'s back to `$e$'s and remove
them. This gives:
\begin{align*}
&\det\omega(\{T_i, Je_i\};\{U_i, Je_i\})\\
=&\det\omega(\{T_1, \ldots, T_9, R_i, Je_i\};\{U_1, U_2, U_3, U_5, U_6,
U_7, U_9, U_{10}, U_{12}, R_i, Je_i\})\\
=& \det(\{R_i\}/\{e_i\})^2\\
&\times  \det\omega(\{T_1, \ldots, T_9, e_i, Je_i\};\{U_1, U_2, U_3, U_5, U_6,
U_7, U_9, U_{10}, U_{12}, e_i, Je_i\})\\
=& \det(\{R_i\}/\{e_i\})^2 \det\omega(\{T_1, \ldots, T_9\};\{U_1, U_2, U_3, U_5, U_6,
U_7, U_9, U_{10}, U_{12}\})
\end{align*}

This remaining $9 \times 9$ determinant has to be done by explicit
calculation. Fortunately the good choice of vectors helps enormously.
The `$T$'s have only three non-zero coordinates (out of 12), the
`$U$'s have only two, and these must overlap if there is to be a
non-zero matrix entry. So a representative non-zero matrix element is
something like
\[ \omega(T^l_v, U^u_w)=\frac{1}{4\pi x^2} [x. (v \times x). (w \times x)] =
\frac{1}{4\pi} [x. v .w]\] where $x$ is whichever of the `$x_i$'s
corresponds to the overlap. (It will be plus or minus one of
$a,b,c,d,e,f$, depending on whether the overlap of
coordinates happens in the first or second of the two non-zero slots of the
`$U$'--vector, respectively.)

Writing down the matrix with rows corresponding to $T_1, T_2, \ldots,
T_9$ and columns corresponding to $U^a_c, U^a_e , U^b_a, U^c_e, U^c_a,
U^d_c, U^e_a, U^e_c, U^f_e$  gives
\[ \frac{1}{4\pi}\left(\begin{matrix} 
0	&0	&0	&[cae]	&0	&0	&0	&0	&0\\
0	&[ace]	&[bca]	&0	&0	&0	&0	&0	&0\cr
[aec]	&0	&[bea]	&0	&[cea]	&0	&0	&0	&0\\
0	&0	&0	&-[cae]	&0	&[dac]	&0	&[eac]	&0\\
0	&0	&0	&0	&0	&0	&[eca]	&0	&0\\
0	&0	&0	&0	&-[cea]	&[dec]	&0	&0	&0\\
0	&0	&0	&0	&0	&0	&0	&-[eac]	&[fae]\\
0	&-[ace]	&0	&0	&0	&0	&-[eca]	&0	&[fce]\\
-[aec]	&0	&0	&0	&0	&0	&0	&0	&0	
\end{matrix}\right) \]
where $[ace]$ is just a shorthand for the vector triple product
$[a.c.e]$. Substituting the relations $b=-a-c, d=c-e, f=a+e$ and extracting the
factor of $[ace]$ gives
\[ \frac{[ace]}{4\pi}\left(\begin{matrix} 
0	&0	&0	&-1	&0	&0	&0	&0	&0	\\
0	&1	&0	&0	&0	&0	&0	&0	&0	\\
-1	&0	&-1	&0	&1	&0	&0	&0	&0	\\
0	&0	&0	&1	&0	&-1	&0	&1	&0	\\
0	&0	&0	&0	&0	&0	&-1	&0	&0	\\
0	&0	&0	&0	&-1	&0	&0	&0	&0	\\
0	&0	&0	&0	&0	&0	&0	&-1	&0	\\
0	&-1	&0	&0	&0	&0	&1	&0	&1	\\
1	&0	&0	&0	&0	&0	&0	&0	&0	
\end{matrix}\right) \]
This matrix has determinant $[ace]^9/(4\pi)^9$. 

The various change-of-basis determinants may be evaluated in terms of
orbit volumes. If we denote by $\sgn(\{X_i\}/\{Y_i\})$ the sign of the
determinant of the appropriate transformation then
\[ \det(\{T_i\}/\{X_i, e_i\}) = \sgn(\{T_i\}/\{X_i,
e_i\})\vol_{B_\tau}\{T_i\}\]
because $\{X_i, e_i\}$ is $B_\tau$--orthonormal. The volume is given by
\[ \vol_{B_\tau}\{T_i\} =
\frac{\vol(G^4\tau)}{\vol_\rho(G)^4}[ace]^4\] by lemma \ref{t:orbvol}
and the fact that the 12 `$T$'s separate into four orthogonal triplets
coming from the Lie algebra elements $a,c,e \in \Rthree$.  Similarly
we have for the `$U$' case:
\[ \det(\{U_i\}/\{Y_i, e_i\}) = \sgn(\{U_i\}/\{Y_i,
e_i\})\vol_{B_\tau}\{U_i\}\] The volume term may be expressed via
lemma \ref{t:rorbvol}. This will involve a product of six terms of the
form $\vol_\rho\{\hat \xi_i\}$: for each choice of $u$, we have to
calculate the area spanned by the vectors $w_1$ and $w_2$ once
projected into the orthogonal complement of $u$. This is just $\vert
[w_1. w_2.u]/u \vert$. Substituting the explicit choices we made gives
\[ \vol_{B_\tau}\{U_i\} =
\frac{\vol(G^6\tau)}{\vol_\rho(G/T)^6}\frac{[ace]^6}{\prod a} \]
where the product on the right hand is simply $abcdef$. Yet another
application of lemma \ref{t:orbvol} yields:
\[\det(\{R_i\}/\{e_i\})^2 =
\left(\frac{\vol(G\tau)}{\vol_\rho(G)}\right)^2[ace]^2 \]
The sign terms here depend on the original choice of orthonormal
bases, which we did not specify. So we do not yet know the actual sign
of $\det(Q)$. Similar terms will appear in the
computation of $\det(P-iQ)$, however, so that the product of the two
terms does not depend on the original choice. So far we have 
\begin{equation}\begin{split}\det(Q)= &-\sgn(\{T_i\}/\{X_i,
e_i\})\sgn(\{U_i\}/\{Y_i,
e_i\})\left(\frac{[ace]}{4\pi}\right)^9 \\
&\quad\times \left(\frac{\vol(G^4\tau)}{\vol_\rho(G)^4}[ace]^4\right)^{-1}\left(\frac{\vol(G^6\tau)}{\vol_\rho(G/T)^6}\frac{[ace]^6}{\prod
a}\right)^{-1} \left(\frac{\vol(G\tau)[ace]}{\vol_\rho(G)}\right)^2
\end{split}\end{equation}

\subsection{Phase of the Hessian}

We work out $\det(P-iQ)$ (defined by $((P-iQ)_{ij} = \overline{H_G(X_i, Y_j)}$)
using similar techniques. We choose slightly different bases in
$T_\tau M$ this time.

\begin{figure}[ht]
\[\vcenter{\hbox{\mbox{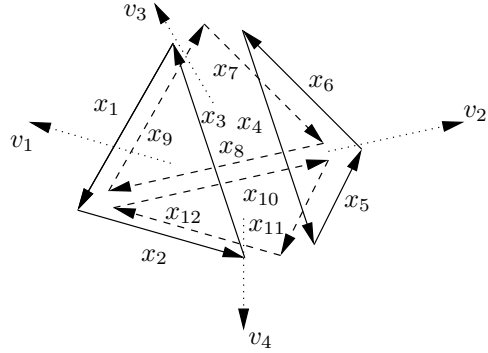}}} \]
\caption{The relevant vectors\label{explode}}\end{figure}

For each face of the tetrahedron $\tau$, numbered by $l$ as
earlier, choose three infinitesimal rotation vectors $T^l_v$ by
letting $v$ be an exterior unit normal vector $v_l$ to the face or one
of the two edges $x_{3l-2}, x_{3l-1}$ of that face. These occur in
clockwise order, so that
\[ x_{3l-2} \times  x_{3l-1} = A_l v_l\]
with $A_l$ twice the area of the $lth$ face. (Of course at the point
$\tau$, we know that each $x_i$ is just one of the vectors
$a,b,c,d,e,f$ or their negatives. However, it is easier to calculate
without substituting these yet.) Pick a set of 12 vectors $U^u_w$
rather as before, except that given an edge $u$, we allow $w$ to be
the exterior unit normal to either of the two faces incident at
$u$. Order the two choices so that the first cross the second points
along $u$ (in fact this corresponds to $v_i$ coming before $v_j$ iff
$i<j$). Figure \ref{explode} shows these where these vectors are in
$\Rthree$, given the tetrahedron $\tau$. We will refer to the chosen
vectors as $U_1', \ldots, U_{12}'$ and $T_1', T_2', \ldots, T_{12}'$.

By a familiar change of basis procedure
\begin{align*}
\det(P-iQ) &= \det H_G(\{Y_i\};\{X_i\}) \\
&=\det H(\{Y_i, e_i\};\{X_i, e_i\}) \\
&=\det(\{T_i'\}/\{X_i, e_i\})^{-1}\det(\{U_i'\}/\{Y_i,
e_i\})^{-1}\det H(\{U_i'\};\{T_i'\}).
\end{align*}

Since we are know the determinant is actually just a phase, we can
throw away any positive real factors appearing during the
computation. For example, the correcting determinants above may
immediately be replace by correcting signs, because the difference (a
volume) is positive. This principle also facilitates the direct
computation of $ \det H(\{T_i'\};\{U_i'\})$ too.

Let us compute sample non-zero elements (once again, most of the
matrix elements will be zero):
\begin{align*}
H(U^u_w, T^l_v) &= \omega(T^l_v, JU^u_w + i U^u_w)\\
&= \frac{1}{4\pi x^3}[x .(v \times x) . (x \times (w \times
x))] + \frac{i}{4\pi x^2}[x .(v \times x).(w \times x)] \\
& = \frac{1}{4\pi} (x(v.w)+i[xvw])\end{align*}
using the earlier notation for the triple product, and with $x$ being
whichever of the `$x_i$'s corresponds to the overlap of the non-zero
coordinates of $U^u_w, T^l_v$. There has been some simplification
because $w.x=0$.

Let us immediately forget about the $4\pi$ factors. If the $i$th and
$j$th faces meet in an edge $u$, oriented along the direction of $v_i
\times v_j$, then the exterior dihedral angle, written $\theta_u$ or
$\theta_{ij}$, is defined (in the range $(0, \pi)$) by
\[ u \sin (\theta_{ij}) = [u.v_i.v_j] \]
\[ \cos(\theta_{ij}) = v_i.v_j. \]

In performing the computation we run across three kinds of non-zero
matrix elements. If $v=v_l$ is normal to the $l$th face, and $w=v_k$
then we obtain $\vert x\vert \e{k}{l} \equiv \vert x \vert
e^{i\theta_{kl}}$. (If $k=l$ it is simply $\vert x \vert$.)  If
instead $v$ is a vector lying in the $l$th face, and $w=v_k$, then we
distinguish according to whether $k=l$ or not. In the case of
equality, we get case we get $0$, $iA_l$ or $-iA_l$ according to
whether $v$ is $x$, its successor, or predecessor in the anticlockwise
cyclic ordering around the face. If $k\neq l$ then we get
$0$, $iA_l \e{k}{l}$ or $iA_l \e{k}{l}$, 
according to the same conditions. 

We can throw out the area factors and the eight powers of $i$ coming
from these second and third cases, and end up with a matrix:
\setcounter{MaxMatrixCols}{12}
\footnotesize
\[\left(
\begin{matrix}
a	&a\e13	&b	&b\e14	&c	&c\e12	&0&0&0&0&0&0\\
0	&0	&-1	&-\e14	&1	&\e12	&0&0&0&0&0&0\\
1	&\e13	&0	&0	&-1	&-\e12	&0&0&0&0&0&0\\
0&0&0&0	&c\e12	&c	&d	&d\e24	&e	&e\e23	&0&0\\
0&0&0&0	&0	&0	&-1	&-\e24	&1	&\e23	&0&0\\
0&0&0&0	&\e12	&1	&0	&0	&-1	&-\e23	&0&0\\
a\e13	&a	&0&0&0&0&0&0	&e\e23	&e	&f	&f\e34	\\
\e13	&1	&0&0&0&0&0&0	&0	&0	&-1	&-\e34	\\
-\e13	&-1	&0&0&0&0&0&0	&\e23	&1	&0	&0	\\
0&0	&b\e14	&b	&0&0	&d\e24	&d	&0&0	&f\e34	&f	\\
0&0	&\e14	&1	&0&0	&-\e24	&-1	&0&0	&0	&0	\\
0&0	&-\e14	&-1	&0&0	&0	&0	&0&0	&\e34	&1	
\end{matrix}
\right)\] 
\normalsize
Easy row operations and then permutation of the rows
reduces the determinant to minus that of the direct sum of the six
$2\times 2$ blocks
\[ \left(\begin{matrix}1&\e{i}{j}\\\e{i}{j}&1\end{matrix}\right)
.\] The determinant of such a block is
\[ -\e{i}{j}.2i \sin(\theta_{ij}) .\]
Discarding the sines, which are positive and so only affect the
modulus of the determinant, we find that its phase is 
\[ e^{i\sum \theta_a}. \]

Let us collect up $\det(P-iQ)\det(Q)$ finally. The annoying sign terms
may be combined into 
\[ \sgn(\{T_i\}/\{T_i'\}) \sgn(\{U_i\}/\{U_i'\}). \]
The first term is a product of four signs arising from orientations of
a three-dimensional vector space, and the second a product of six
arising from two-dimensional spaces. All these signs are positive
(an easy check). 

Thus we have for the Hessian: 
\begin{equation}\label{e:hess} \begin{split}
\Hess_{[\tau]}(\psi) &=
-i(2\pi)^{18}2^9 e^{i\sum
\theta_a}\left(\frac{[ace]}{4\pi}\right)^9\\
&\qquad  \times 
\left(\frac{\vol(G^4\tau)}{\vol_\rho(G)^4}[ace]^4\right)^{-1}
\left(\frac{\vol(G^6\tau)}{\vol_\rho(G/T)^6}\frac{[ace]^6}{\prod
a}\right)^{-1}
\left(\frac{[ace]}{\vol_\rho(G)}\right)^2\\
& = -ie^{i\sum
\theta_a}(2\pi)^9[ace]\left(\prod
a\right)\left(\frac{\vol_\rho(G)^2\vol_\rho(G/T)^6}{\vol(G^4\tau)\vol(G^6\tau)}
\right)\end{split}\end{equation}

Looking at the argument again, it is easy to see that the Hessian at
$[\tau']$ is the complex conjugate of this one.

\subsection{The overall phase of the integrand}\label{phase}

We need to account for the 0--order contributions $\psi([\tau']),
\psi([\tau])$ of the integrand at the two critical points. Since the
two sections being paired were fixed to have norm 1 along their
critical regions, these 0--order contributions also have modulus 1. We
start by calculating the phase {\em difference}.

The chosen lifts $\tau, \tau '$ of these points lie on the slice
$\mu=0$ inside $M$. Since this slice is just a product of four
``spaces of triangles'', each of which is a principal $G$--space,
($G=SO(3)$) there is a unique element of $G^4$ which translates $\tau$
to $\tau '$. In fact it is easy to describe such an element $g=(g_1,
g_2, g_3, g_4)$ explicitly. The element $g_1$ must rotate the triangle
$(a,b,c)$ (the projection of $\tau$ into the first three sphere
factors of $M$) to its negative $(-a,-b,-c)$. Therefore it is the
rotation of $\pi$ about the normal to the triangle's plane. The other
$g_i$ are similarly half-turns normal to their respective triangular
faces.

We must compare the values of the pointwise pairing $\tilde s_\mu,
\tilde s_\nu$ at $\tau, \tau '$. Define for each face a lift $\tilde
g_i$ of $g_i$ into $SU(2)$ by lifting the path of anticlockwise
rotations from $0$ to $\pi$. Together these form $\tilde g \in
SU(2)^4$. Since $\tilde s_\mu$ is $SU(2)^4$--invariant:
\[ \tilde s_\mu (\tau') = \tilde s_\mu (\tilde g\tau) =
\tilde g \tilde g^{-1}(\tilde s_\mu(\tilde g \tau))= \tilde g ((\tilde
g^{-1}\tilde s_\mu)(\tau)) = \tilde g (\tilde s_\mu(\tau)) \] By
contrast, $\tilde s_\nu$ is not $SU(2)^4$--invariant, though it is
$SU(2)^6$--invariant. We can write an equation like the above but we
need to know what $(\tilde g^{-1}\tilde s_\nu)$ is to perform the last step.
Now $\tilde s_\nu$ is a sextuple tensor product, and we can study the
action of $\tilde g^{-1}$ on it by looking at the action on the six factors
individually:
\[\tilde  g^{-1}\tilde s_\nu = (\tilde g_1^{-1}, \tilde
g_3^{-1})s^{aa} \otimes (\tilde g_1^{-1}, \tilde g_4^{-1}) s^{bb}
\otimes \cdots \otimes (\tilde g_3^{-1}, \tilde g_4^{-1})s^{ff}\]
Using the diagonal invariance of each section of the form $s^{aa}$, we
have identities like \[ (\tilde g_1^{-1}, \tilde g_3^{-1})s^{aa} = (1,
\tilde g_3^{-1}\tilde g_1)s^{aa}.\] Now $\tilde g_1, \tilde g_3$ are
lifts of rotations through $\pi$ about directions normal to the two
faces of the tetrahedron meeting at side $a$, so the composite $\tilde
g_3^{-1}\tilde g_1$ is a lift of an anticlockwise rotation through an
angle equal to twice the exterior dihedral angle $\theta_a$ about the
vector $a$, but it is slightly tricky to decide {\em which} lift. Let
us denote by $\tilde r_a$ the lift of the path of anticlockwise
rotations from $0$ to $2\theta_a$, and write $\tilde g_3^{-1}\tilde
g_1= \delta_a \tilde r_a$, for some $\delta_a=\pm 1$. Then
\begin{align*}
((\tilde g_1^{-1}, \tilde g_3^{-1})s^{aa})(a, -a) & = (1,
\delta_a\tilde r_a)(s^{aa}((1,\delta_a\tilde r_a^{-1})(a,-a)))\\
& = (1,
\delta_a\tilde r_a)(s^{aa}(a,-a)).\end{align*} The action of $\tilde r_a$ on the
fibre of the bundle $\LL^{\otimes a} \ra \PP^1$ at $-a$ is
multiplication by $e^{-ia\theta_a}$ (remember that it acts as
$e^{i\theta_a}$ on the fibre of the tangent bundle $\LL^{\otimes 2}$
at $a$). The sign $\delta_a$ acts as its $a$th power $\delta_a^a$.  So,
using the invariance of the hermitian form, 
\begin{align*} \lan \tilde s_\mu (\tau'), \tilde s_\nu(\tau') \ran & =  \lan \tilde
g (\tilde s_\mu(\tau)), \tilde g (\left(\prod \delta_a^a\right)
e^{-i\sum a\theta_a}s_\nu (\tau) )\ran\\
& = \left(\prod
\delta_a^a\right)e^{i\sum a\theta_a}\lan \tilde s_\mu(\tau), \tilde s_\nu
(\tau)\ran. \end{align*} In fact this identity is independent of the choice of
lifts $\tilde g_i$. For example, changing the lift of $\tilde g_1$
negates $\delta_a, \delta_b, \delta_c$, changing the right-hand side
by $(-1)^{a+b+c}$, which is $+1$ because of the parity condition on
$a,b,c$. Further, if we imagine varying the dihedral angles of the
tetrahedron in the range $(0,\pi)$, all our chosen lifts are
continuous and so we may evaluate the sign $\left(\prod
\delta_a^a\right)$ by deformation to a flat one, for example where
$\theta_a, \theta_b, \theta_c$ are $\pi$ and the others $0$. In this
case, $\tilde g_2=\tilde g_3 =\tilde g_4 = - \tilde g_1$, and so all
the $\delta_a$ turn out positive. Hence $\psi([\tau'])=e^{ik\sum a
\theta_a}\psi([\tau])$. Because the $6j$--symbol itself is {\em real},
the two values of $\psi$ must be conjugate. Therefore
$\psi([\tau'])=\pm e^{\frac12ik\sum a \theta_a}$, but we still have
an annoying sign ambiguity.

There are really three separate sign problems: how the sign depends on
$k$ (for fixed $a,b, \ldots, f$); how it varies as we alter $a,b,
\ldots, f$; and one overall choice of sign. One step is easy: the
phase conventions on the sections used in the pairing implied for
example that $s^{ka, ka} = (s^{aa})^{\otimes k}$ and that the
integrand was a $k$th power of another function, the sign above must
actually be $(-1)^k$ or or $1$.

To do better requires a frustrating amount of work, which will only be
sketched here. Recall that the signs in definition \ref{6jdef} were
fixed by writing down explicit polynomial representatives of the
trilinear and bilinear invariants. If we remove a suitable branch cut
from each copy of the sphere in $M$, leaving a contractible manifold,
we can extend this definition to allow real values of the variables
$a,b, \ldots, f$, and extend $\psi([\tau])$ to a {\em real-analytic}
function of these variables (at least locally). It is obtained by
pairing two holomorphic sections of a trivial bundle with a hermitian
structure which still satisfies the quantization formula
\eqref{e:quant}. There is another way of computing the phase
difference above, based on choosing two paths $\tau \ra \tau'$ in $M$,
one on which $\mu=0$ and one on which $\nu=0$, and computing the
holonomy around the resulting loop. This can be done by computing the
symplectic area of a bounded disc. To carry this out appropriately one
must be very careful with {\em which} disc: once the symplectic form
no longer has integral periods on $H_2(M)$, this matters. Further, the
most obvious paths and disc intersect the branch cuts, so one must
account for this too. Ultimately one obtains an analytic expression
\[ \psi([\tau'])=\pm e^{\frac12ik\sum a \theta_a+i\sum a\pi}\]
where the analyticity restricts this sign to a single {\em overall}
ambiguity. (Note that at integral values of the lengths, we can see
the sign $(-1)^{\sum a}$ appearing.) One could compute this
sign using a single example, but as the reader will judge from this
terse paragraph, the author is so bored with fixing signs that he no
longer cares to! The experimental evidence in \cite{PR} confirms that
the sign is positive.

Therefore
\begin{equation}\label{e:0order} \psi([\tau']=(-1)^{\sum
a}e^{\frac12ik\sum a \theta_a} \quad \hbox{and} \quad
\psi([\tau])=(-1)^{\sum a}e^{-\frac12ik\sum a\theta_a}.\end{equation}

\subsection{Putting it all together}

We combine the original integral definition \eqref{e:main} with the
asymptotic normalisation factors \eqref{e:gnorm}, \eqref{e:gnorm2},
the reduction \eqref{e:red}, the stationary phase evaluation
\eqref{e:stat} incorporating the Hessian \eqref{e:hess} and 0--order
terms \eqref{e:0order}. 
\begin{equation*}\begin{split}
\ksixj &\sim \left(\prod\sqrt{ka+1}\right)
\left(\frac{k}{2}\right)^{-6}(\vol(\mu^{-1}(0))\vol(\nu^{-1}(0)))^{-\frac12}
 k^9
\left(\frac{k}{2}\right)^{\frac32}\\
&\qquad\times \left(\frac{2\pi}{k}\right)^{9}\left\{\frac{e^{-\frac12ik\sum
a \theta_a}}{\sqrt{-\Hess_{[\tau]}(\psi)}} +\frac{e^{\frac12ik\sum
a
\theta_a}}{\sqrt{-\Hess_{[\tau']}(\psi)}}\right\}.\end{split}\end{equation*}
The
terms from the Hessian and the normalisation involving
\[\vol(\mu^{-1}(0))=\vol(G^4\tau)=\vol(G^4\tau')\ \hbox{and}\ 
\vol(\nu^{-1}(0))=\vol(G^6\tau)=\vol(G^6\tau')\] cancel. The
normalisation factor $(\prod (ka+1))^{\frac12}$ cancels with the term in the
Hessian involving $\prod a$, contributing asymptotically simply
 $k^3$.  What remains is
\[(2\pi)^{\frac92}2^{\frac{11}{2}}k^{-\frac32}[ace]^{-\frac12}\vol_\rho(G)^{-1}\vol_\rho(G/T)^{-3}\cos{\left\{ \sum (ka+1)
\frac{\theta_a}{2} + \frac{\pi}{4}\right\}}.\] Substituting in the volumes
$8\pi^2$ and $4\pi$ of $G$ and $G/T$ gives
\[ \ksixj \sim \sqrt{\frac{2}{3\pi k^3V}} \cos{\left\{ \sum (ka+1)
\frac{\theta_a}{2} + \frac{\pi}{4}\right\}} \]
where $V=\frac16[ace]$ is the (scaling-independent) volume of $\tau$. This
completes the proof of the theorem.


\section{Further geometrical remarks}

\subsection{Comparison with the Ponzano--Regge formula}

It is important to note that the formula \eqref{e:form} is {\em not}
the same as the original Ponzano--Regge formula. There are two main
differences, apart from the trivial fact that they label their
representations by half-integers instead of integers.

Their claim, in our integer-labelling notation, is that for large
$a,b,c,d,e,f$:
\begin{equation}\label{e:formpr} \sixj \approx \begin{cases}{\displaystyle\sqrt{\frac{2}{3\pi V'}}\cos{\left\{ \sum (a+1)\frac{\theta_a'}{2} + \frac{\pi}{4}\right\}}} &\hbox{{if $\tau'$ is
Euclidean,}} \\ \hbox{{exponentially decaying}}&\hbox{{if $\tau'$ is
Minkowskian,}}\end{cases} \end{equation} where $\tau'$ is a
tetrahedron whose edges are $a+1, b+1, \ldots, f+1$ and whose dihedral
angles $\theta_a'$ and volume $V'$ are therefore slightly different
from those of our $\tau$. This difference is worrying, as it is quite
possible to find sextuples of integers such that $\tau'$ is Euclidean
yet $\tau$ is Minkowskian, in which case the formulae seem to
conflict: is the $6j$--symbol exponentially or polynomially decaying in
this case?

The second difference explains this. The Ponzano--Regge formula
\eqref{e:formpr} is only claimed as an {\em approximation} for large
irreducibles, rather than an asymptotic expansion in a strict sense as
in theorem \ref{t:main}. Therefore the only meaningful comparison
between the two formulae is to examine how their function behaves as
we rescale $a,b,c,d,e,f$ by $k \ra \infty$ in the precise sense of our
theorem.

Although for small $k$ it is possible that $\tau'$ might be Euclidean
when $\tau$ is not, eventually the shift in edge-lengths becomes
insignificant and either both are Euclidean or neither is. Therefore
there is no inconsistency between cases in the two formulations.

As for comparing the actual formulae, the asymptotic behaviour of the
Ponzano--Regge function is 
\begin{equation}\label{e:formpr2} {\displaystyle\sqrt{\frac{2}{3\pi k^3V}}\cos{\left\{ \sum (ka+1)
\frac{\theta_a'}{2} + \frac{\pi}{4}\right\}}}, \end{equation} because
$V$ and $V'$ agree to leading order in $k$. The only problem is the
dihedral angles relating to slightly different tetrahedra. Fortunately
we may easily show that
\[ e^{ik\sum (a+1)
\theta_a'} \sim e^{ik\sum (a+1) \theta_a}\] by applying the {\em
Schl\"afli identity} (see Milnor \cite{M2}), which says that the
differential form $\sum a d\theta_a$ vanishes identically on the space
of Euclidean tetrahedra. Therefore there is no inconsistency.

\begin{rem} The case of a flat tetrahedron is not covered by either
formula. \end{rem}

\subsection{Regge symmetry and scissors congruence}

Suppose one picks out a pair of opposite sides of the tetrahedron
denoting the $6j$--symbol (as in figure \ref{figtet}), say $a,d$. Let
$s$ be half the sum of the other four labels (twice their
average). Define:
\begin{equation}\label{f:regge}
\begin{matrix} a'=a&\qquad&b'=s-b\\
d'=d&\qquad&c'=s-c\\
&\qquad&e'=s-e\\
&\qquad&f'=s-f
\end{matrix}
\end{equation}
Regge discovered that the $6j$--symbols are invariant under this
algebraic operation (the easiest way to see this is to look at the
generating function for $6j$--symbols, \cite{V}):
\[ \sixj = \sixjj\]
We can also consider this as a {\em geometric} operation on a
tetrahedron, altering its side lengths according to the above
scheme. It is not meant to be obvious that the result of applying
this to a Euclidean tetrahedron will return a Euclidean one!

Regge and Ponzano considered the effect of this symmetry on the
geometrical quantities occurring in their asymptotic formula, mainly
as another check on its plausibility. They discovered that the volume
and phase term associated to a Euclidean tetrahedron are indeed
exactly invariant. This is amazing, given that it would be consistent
with their appearance in an {\em asymptotic} expansion for them to
change, but by lower-order contributions.

Let us reconsider this surprising geometric symmetry.  First note that
the symmetry is an involution: if one thinks geometrically, it
corresponds to reflecting the lengths of the four chosen sides about
their common average. These involutions, together with the tetrahedral
symmetries, form a group of 144 symmetries of the $6j$--symbol,
isomorphic to $S_4 \times S_3$ (see \cite{V}).

V.~G.~Turaev pointed out to me that the term $\sum l_i \theta_i$ in
the phase part of the formula \eqref{e:form} for a tetrahedron $\tau$
is reminiscent of the Dehn invariant $\delta(\tau)$. Actually it would
be fairer to say that it is the ``Hadwiger measure'' (or ``Steiner
measure'') $\mu_1(\tau)$.  Both invariants are connected with problems
of equidissection of three-dimensional polyhedra. 

Two polyhedra are {\em scissors congruent} if one may be dissected
into finitely-many subpolyhedra which may be reassembled to form the
other. (Hilbert's third problem was to determine whether
three-dimensional polyhedra with equal volumes were, as is the case in
two dimensions, scissors-congruent. Dehn used his invariant to solve
this problem in the negative.)

The modern way of looking at the problem is to define a Grothendieck
group of polyhedra $\Pol$. We take $\Z$--linear combinations of
polyhedra in $\Rthree$ with the relations:
\begin{align}
P \cup Q &= P + Q - P\cap Q \\ \label{e:deg}P&=0 \qquad \hbox{if $P$ is
degenerate}\\ P&=Q \qquad \hbox{if $P$, $Q$ are congruent}
\end{align}
Volume is an obvious homomorphism $\Pol \ra \R$. The Dehn invariant is
a less obvious one $\Pol \ra \R \otimes_{\Z} (\R/\pi \Z)$, defined for
a polyhedron by summing, over its edges, their lengths tensor dihedral
angles:
\[ \delta(P) = \sum l_i \otimes \theta_i\]
Sydler proved that these two invariants suffice to {\em classify}
polyhedra up to scissors-congruence: two such are scissors-congruent
{\em if and only if} they have the same volume and Dehn invariant. See
Cartier \cite{C} for more details.

If we look for homomorphisms $\Pol \ra \R$ which are {\em continuous}
under small perturbations of vertices of a polyhedron, then volume is
the only one (up to scaling). However, if we remove the condition
\eqref{e:deg} on degenerate polyhedra from the axioms defining $\Pol$,
there is a four-dimensional vector space of continuous homomorphisms,
spanned by the following {\em Hadwiger measures} (picked out as
eigenvectors under dilation):
\begin{align}
\mu_3(P) &= \vol(P)\\
\mu_2(P)&= \textstyle{\frac12} \area(\partial P)\\
\mu_1(P) &= \textstyle{\sum} l_i \theta_i\\
\mu_0(P) &= \chi(P) \quad \hbox{(the Euler characteristic)}
\end{align}
See Milnor \cite{M} or Klain and Rota \cite{KR} for more on these
beautiful functions.

The relationship with the Regge symmetry of tetrahedra is as follows:

\begin{theorem}\label{t:Dehn}
The Regge symmetry \eqref{f:regge} takes Euclidean tetrahedra to
Euclidean tetrahedra, preserving volume, Dehn invariant and Hadwiger
measure $\mu_1$. {\rm (}Remark: simple examples show that Regge symmetry
does not preserve the surface area measure $\mu_2$.{\rm )}
\end{theorem}
\begin{proof}
The tour-de-brute-force of trigonometry in appendices B and D of
\cite{PR} contains all the calculations necessary to prove this. They
demonstrate that under the Regge symmetry, which is a rational linear
transformation $A$ of the six edge lengths, the dihedral angles also
transform according to $A$. The orthogonality of this matrix and the
fact that we may view the Dehn invariant as being in $\R \otimes_{\Q}
(\R/\pi \Z) \equiv \R \otimes_{\Z} (\R/\pi \Z)$ demonstrate its
invariance, as well as that of $\mu_1$. The volume is checked by
straightforward calculation using the Cayley determinant.
\end{proof}

\begin{coroll}
The orbit under the group of 144 symmetries of a generic tetrahedron
consists of {\em twelve} distinct congruence classes of tetrahedra,
all of which are scissors-congruent to one another.
\end{coroll}

\begin{rem}
The fact that a tetrahedron is scissors-congruent to its mirror-image
was proved by Gerling in 1844 (see Neumann \cite{N}), using
perpendicular barycentric subdivision about the circumcentre. One
would expect that for the Regge symmetry, which is also a ``generic''
scissors congruence (as opposed to a ``random'' coincidence of volume
and Dehn invariant for two specific tetrahedra), a similar general
construction might be given. What is it? 
\end{rem}

\subsection{Further questions}

1.\qua Ponzano and Regge give an explicit formula for the exponential
decay of the $6j$--symbol, in the case when no Euclidean tetrahedron
exists. Amazingly, it is an analytic continuation of the main formula,
incorporating the volume of the {\em Minkowskian} tetrahedron which
exists instead, and with the oscillatory phase term converted into a
decaying hyperbolic function. Can this be extracted from a similar
procedure? 

2.\qua  Can similar geometrically-meaningful formulae be obtained for
general spin networks, the so-called $3nj$--symbols?

3.\qua  The calculation in this paper is comparatively crude, since it
computes a pairing of 12--linear invariants when one could really do
with a pairing of 4--linear invariants (see the remark of section
\ref{4def}). The space whose quantization gives quadrilinear
invariants is 2--dimensional, in fact a sphere with a non-standard
K\"ahler structure. Can one work directly on this space instead?
(Possibly any advantage in dimensional reduction is lost when one
needs to do explicit calculations, which end up like the ones here).

4.\qua  Can similar formulae be obtained for other groups, and does their
associated geometry have any physical meaning? The $6j$--symbols are
scalars only for multiplicity-free groups such as $SU(2)$. In general
they live in the tensor product of four trilinear invariant spaces, in
which one would need preferred bases.

5.\qua  Can one obtain similar formulae for quantum $6j$--symbols, which
arise as pairings in the quantization of moduli spaces of flat
connections on the 4--punctured sphere?

\np

\end{document}